%% file: angmom34-arxiv.tex
\documentclass{rspublic}

\usepackage{url}
\usepackage{natbib}
\usepackage{latexsym}
\usepackage{amsmath}
\usepackage{amstext,amsfonts,epsf,epsfig,pifont}
\usepackage{graphics,graphicx,bm,amsmath,multicol}

\newlength{\intwidth}

\def\Xint#1{\mathchoice
{\XXint\displaystyle\textstyle{#1}}%
{\XXint\textstyle\scriptstyle{#1}}%
{\XXint\scriptstyle\scriptscriptstyle{#1}}%
{\XXint\scriptscriptstyle\scriptscriptstyle{#1}}%
\!\int}
\def\XXint#1#2#3{{\setbox0=\hbox{$#1{#2#3}{\int}$}
\vcenter{\hbox{$#2#3$}}\kern-.5\wd0}}

\def\dashint{\Xint-}

\newcommand{\cross}{\ensuremath{{\mathrm{cr}}}}
\newcommand{\angmom}{\ensuremath{\mathbf{m}}}

\newcommand{\mherm}{\ensuremath{\mathrm{herm}}}
\input{texheader}

\begin{document}
\title[The unity of moments]{The unity of instantaneous spectral moments and physical moments}

\author[J. M. Lilly]{Jonathan M. Lilly}

\affiliation{NorthWest Research Associates, PO Box 3027,
Bellevue, WA 98009, USA}

\label{firstpage}

\maketitle
\begin{abstract}{Instantaneous frequency, instantaneous bandwidth, nonstationary signal analysis, trivariate signal, angular momentum, circulation.}
A modulated oscillation in two or three dimensions can be represented as the trajectory traced out in space by a particle orbiting an ellipse, the properties of which vary as a function of time.   Generalizing ideas from signal analysis, the signal variability can be described in terms of kinematic quantities, the {\em instantaneous moments},  that formalize our intuitive notions of time-varying frequency and amplitude.   On the other hand, if we observed an ellipse evolving in space we would seek to describe it in terms of its  physical moments,  such as  angular momentum,  moment of inertia, etc.   The main result of this paper is to show that the two sets of moments are identical.    Most significantly,  an essential physical quantity---the circulation---is the same as the product of the two most important kinematic quantities, the instantaneous frequency and the squared instantaneous amplitude.   In addition to providing a rich set of geometric tools for the analysis of nonstationary oscillations in two or three dimensions, this result also has implications for the practical problem of inferring physical ellipse parameters from the trajectory of a single particle on the ellipse periphery, as is frequently encountered in the study of vortex motions.   
\end{abstract}

\section{Introduction}

A powerful means emerging from the signal processing community for the analysis of nonstationary signals is the what may be termed the {\em method of instantaneous moments}, introduced by \citet{gabor46-piee} and extended by many authors \citep[e.g.][]{ville48-cet,bedrosian63-ire,vakman77-spu,cohen89-ieee,boashash92a-ieee,cohen,vakman96-itsp,picinbono97-itsp,loughlin00-ispl}.    The instantaneous moments are functions of time, derived from the signal itself, that offer {\em time-varying} generalizations of our standard intuitive concepts---rooted in Fourier analysis, hence constant in time---of signal amplitude, frequency, and bandwidth.  Since the instantaneous moments time-average to the corresponding Fourier-domain moments of the signal's spectrum, these quantities provide a direct and compelling link between time variability and frequency-domain structure. This method had its genesis when \citet{gabor46-piee} brought ideas from quantum mechanics to bear on outstanding questions in signal processing questions,  formalizing the analysis of the information content of signals in a way that had not previously been possible.    


In recent years, as the intuitive content of the instantaneous moments has become more clear---thanks to work such as \cite{vakman96-itsp}, \citet{cohen89-ieee}, \citet{cohen}, and \citet{loughlin00-ispl}---the virtue of this method for the analysis of time-varying signals in observational data has become increasingly apparent.    In both geophysical \citep{taner79-geo,rene86-geo,diallo06-geo,lilly11-itsp} and  oceanographic  \citep{lilly06-npg,lilly11-grl} applications, ideas based upon the method of instantaneous moments are proving to be remarkably informative for studying oscillatory processes that evolve with time.   Thus ideas from physics brought to, and evolved in, the signal processing community are now moving back into the physical sciences, providing a much-needed mathematical foundation for the analysis of nonstationary signals.


A full appreciation of the power and potential of the method of instantaneous moments to the analysis of physical signals requires another generalization.   This method has been developed for a univariate signal, i.e. a scalar-valued function of time $x(t)$, whereas in the physical sciences, it is more frequently the case that the signal of interest is a vector $\bx(t)$ in two or three dimensions.  This is so because oscillations are typically generated by waves or wavelike processes in physical space.   A step in this direction was recently taken by \citet{lilly10-itsp}, who derived scalar-valued functions of time, called the {\em joint instantaneous moments}, that play the same roles for a multivariate signal $\bx(t)$ that the standard instantaneous moments play for a univariate signal $x(t)$.    The joint instantaneous moments have been shown to have informative expressions in terms of the geometry of the time-varying ellipse traced out by the signal in two  \citep{lilly10-itsp} or three  \citep{lilly11-itsp} dimensions.  Recovery of multivariate oscillations from potentially noisy time series, together with estimation of the joint instantaneous moments,  can then be accomplished by a multivariate generalization \citep{lilly12-itsp} of the wavelet ridge analysis method of \citet{delprat92-itit}.

As  with the instantaneous moments of univariate signal, the joint instantaneous moments introduced by \citet{lilly10-itsp} constitute a scalar-valued function of time at each order, corresponding to the orders of frequency-domain moments of the signal's spectrum.   Yet since the spectrum of a vector-valued signal is matrix-valued, we infer that the complete generalizations of instantaneous moments to such a signal should also be matrix-valued functions of time, and therefore the joint instantaneous moments must therefore only contain a fraction of the available information.  Such matrix-valued instantaneous moments have not yet been defined, and consequently the meaning of the additional information they contain has not yet been explored.     In addition to achieving a more complete understanding of signal variability, there is a second compelling reason to examine in detail the instantaneous moment matrices:  to close the apparent gap between the instantaneous spectral moments on the one hand, and familiar physical moments on the other.  The instantaneous moments could be aptly described as kinematic quantities, in that they make no reference to physical properties.  Yet if we have a particle moving in space we would immediately calculate its angular momentum as a function of time, which is a physical moment.   The link between physical moments and kinematic moments, if any, is not at all apparent.  It is essential to establish this relationship if the method of instantaneous moments is to be used to illuminate the analysis of physical phenomena.  


The purpose of this paper is to fully generalize the instantaneous spectral moments to the trivariate case,  and to establish the connection between these quantities and standard physical quantities, building on work on the scalar-valued joint instantaneous moments for the bivariate and trivariate cases by  \citet{lilly10-itsp} and  \citet{lilly11-itsp} respectively.   It is seen that the connection is surprisingly deep: the matrix-valued instantaneous spectral moments, and the physical moments of an elliptical ring of particles that is implicitly assigned to the time series by the foundational step in the analysis method, {\em are identical}.   Most significantly,  arguably the most important physical moment of the conjectured system---its {\em circulation} or average vorticity---is found to be identical to the product of the two most important kinematic moments, the instantaneous frequency and squared instantaneous amplitude.    This has significant implications for the recovery of vortex properties from particle trajectories, a practical problem that is encountered in oceanography \citep{lilly06-npg,lilly11-grl} and in fluid dynamics more generally. 

Fundamentals of the method of instantaneous moments for multivariate signals, together with necessary preliminary material from the signal processing literature, are presented in \S \ref{section:fundamentals}.  The main result---the unity of moments---is stated without proof in \S \ref{section:unity}, and its implications are discussed.   The proof of the unity of moments is given in \S \ref{section:proof}.  In \S \ref{section:moments},  the interpretations of the instantaneous moments are investigated in more detail.  For reasons of space, this paper focuses on theoretical as opposed to practical issues.   The connection to practical applications is briefly discussed along with the conclusions in \S \ref{section:conclusions}.

\section{Fundamentals}\label{section:fundamentals}

The variability of a trivariate signal can be usefully described in terms of a particle orbiting a time-varying ellipse,  building on a powerful method from the signal processing literature for linking time-domain variability to frequency-domain structure.   In the trivariate case, frequency-domain structure is captured by the $3\times 3$ spectral matrix.  Recent work has established the relationship between the ellipse geometry and the trace of this matrix, but the connection to the remaining terms has not yet been investigated.

\subsection{An elliptical ring of particiles}\label{mvim}

A real-valued signal $\bx(t)=\begin{bmatrix} x(t)  \,\,\, y(t) \,\,\, z(t) \end{bmatrix}^T$  is observed in three dimensions, which for clarity is imagined to be the trajectory traced out by a hypothetical particle in physical space.  The  signal is regarded as deterministic and zero-mean, and is assumed to have been observed over its entire duration and to be of finite energy, i.e. $\int_{-\infty}^\infty  \|\bx(t)\|^2  \rd t < \infty$.   Here $\|\bz\|^2\equiv\bz^H\bz$ is the vector norm of some possibly complex-valued vector  $\bz$, with the superscript ``$H$'' denoting the Hermitian transpose.  There is reason to believe that the trivariate signal $\bx(t)$ is oscillatory in nature, but with properties that may change in time.   Our goal is to analyse $\bx(t)$ in such a way as to illuminate its structure. 

\begin{figure}[t]
        \noindent\begin{center}\includegraphics[width=3.5in,angle=0]{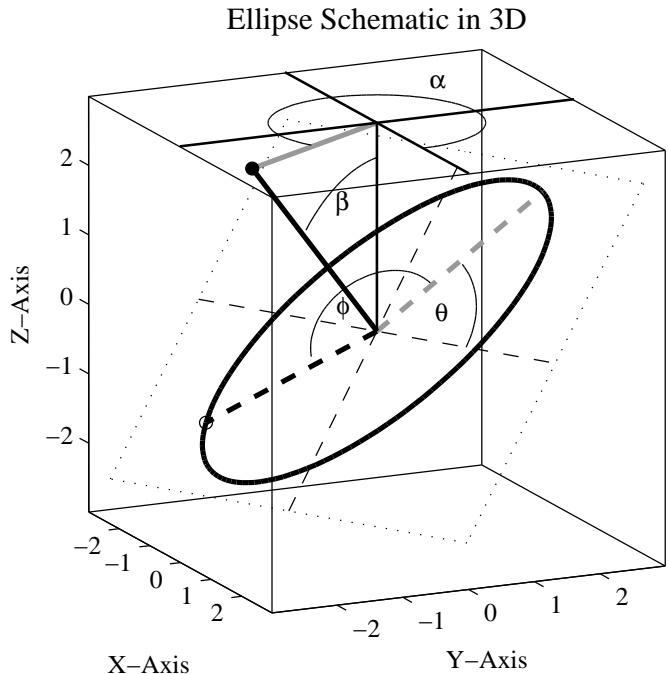}\end{center}
        \caption{\footnotesize Schematic of an elliptical ring of particles in three dimensions.  The ellipse has semi-major axis $a=3$, semi-minor axis $b=3/2$, precession angle $\theta=\pi/3$, zenith angle $\beta=\pi/4$, and azimuth angle $\alpha=\pi/6$.  The phase increases from an initial value at $\phi=5\pi/6$, tracing out the heavy back curve through one full cycle, during which time all other ellipse parameters are constant.  The plane of the ellipse is indicated by the dotted lines, with the original $x$- and $y$-axes marked by thin dashed lines.  A heavy dashed black line shows position of the ``particle'' at the initial time, while a heavy gray dashed line marks the ellipse semi-major axis. A heavy solid black line, with a black circle at its end, is the normal vector to the plane of ellipse; the projection of this vector onto the $x$--$y$ plane is shown with a heavy solid gray line.
        }\label{angmom-schematic}
\end{figure}

The kinematic model of a time-varying ellipse will form the starting point. The evolving position of an elliptical ring of particles, constrained to remain coplanar and elliptical in shape but otherwise free to evolve in three dimensions about its centre of mass, may be written as
\begin{equation}
\bx(t,\varphi) \equiv 
 \bQ_\bx(t)
        \begin{bmatrix} a(t) & 0 \\
 0 & b(t)  \\ 0 & 0
 \end{bmatrix}\begin{bmatrix} \cos \phi(t) & -\sin\phi(t) \\
\sin \phi(t) & \cos \phi(t)
   \end{bmatrix}  \begin{bmatrix} \cos\varphi
\\ \sin \varphi     \end{bmatrix}  \label{ellipsemodel}
\end{equation}
where $\varphi$ is a particle label with $-\pi\le\varphi<\pi$.   Here $a(t)$ and $b(t)$ are the time-varying ellipse semi-major and semi-minor axis lengths with $a(t) \ge b(t)\ge 0$, and the {\em orbital phase} $\phi(t)$ specifies a time-varying shift of particle locations around the ellipse periphery.  $\bQ_\bx(t)$ is a $3\times 3$ rotation matrix written as
\begin{equation}\bQ_\bx(t)\equiv\bJ_3(\alpha(t))\, \bJ_1(\beta(t))\, \bJ_3(\theta(t))\end{equation}
and represents a time-dependent rotation in the so-called ``Z-X-Z'' form, where
\begin{equation}
\bJ_1(\beta) \equiv
    \begin{bmatrix}1& 0& 0\\
   0 & \cos\beta & -\sin\beta \\
   0 & \sin\beta  & \cos\beta
   \end{bmatrix},\quad \quad
   \bJ_3(\alpha) \equiv
    \begin{bmatrix}\cos\alpha&-\sin\alpha & 0\\
   \sin\alpha & \cos\alpha& 0\\
   0 & 0 & 1
   \end{bmatrix}
\end{equation}
are rotation matrices about the $x$ and $z$ axes, respectively.   A sketch of the ellipse geometry is shown in figure~\ref{angmom-schematic}.   In this rotation, $\theta(t)$ sets the orientation of the major axis of the ellipse in a two-dimensional plane, while $\beta(t)$ and $\alpha(t)$ are the zenith and azimuth angles of the normal to the plane containing the ellipse.  From right to left, the kinematic model (\ref{ellipsemodel}) states that a particle labeled by phase $\varphi$ is shifted $\phi(t)$ radians counterclockwise around the unit circle in the $x$--$y$ plane, which is then deformed along the $x$-axis and $y$-axis to generate an ellipse, and which is finally subjected to a rotation in three dimensions.    As $t$ varies with fixed $\varphi$, $\bx(t,\varphi)$ traces out the trajectory in three-dimensional space of the particle labeled by $\varphi$.

\normalsize

This kinematic model has considerable dynamical relevance.  Setting $\alpha=\beta=0$ leads to a two-dimensional version of  (\ref{ellipsemodel}) that emerges frequently in fluid dynamics, encompassing the oscillating elliptical vortex solution to the rotating shallow water equations examined by \citet{young86-jfm} and \citet{holm91-jfm} as well as the archetypal Kirchhoff and \citet{kida81-jpsj} family of two-dimensional vortex solutions.  In fact,  (\ref{ellipsemodel}) with $\alpha=\beta=0$ is identical with equation (2.6) of \citet{holm91-jfm} after substituting his (4.2).  In that work, a set of evolution equations for the ellipse parameters is derived in which periodic motion arises as a consequence of physical conservation laws.  For this and other vortex solutions, the kinematic model (\ref{ellipsemodel}) would be seen as generating a family of trajectories on the left-hand side based on prescribed or physically determined ellipse parameters on the right-hand side.  The primary concern here will be the opposite, namely,  {\em given} a signal we will seek to {\em assign} or {\em infer} a set of ellipse parameters in some meaningful way, thus moving from the left-hand side of  (\ref{ellipsemodel})  to the right-hand side.

\subsection{The canonical ellipse parameters}\label{section:canonical}

An infinite family of ellipse parameters can generate the same physical signal.   The method for assigning a  {\em unique} set of physically meaningful time-varying ellipse parameters to a trivariate signal $\bx(t)$  beings with the construction of the {\em analytic signal}, the starting point for the method of instantaneous moments.  Here we follow the recent multivariate development in \citet{lilly10-itsp} and \citet{lilly11-itsp}; see \cite{boashash92a-ieee} and \cite{picinbono97-itsp} and references therein for background.   With $\calH \bx(t)\equiv \frac{1}{\pi}\,\dashint_{-\infty}^\infty\frac{\bx(u)}{t-u}\,\rd u$ being the Hilbert transform of $\bx(t)$, where  ``$\dashint$'' is the Cauchy principal value integral, the {\em analytic part} of $\bx(t)$ is given by
\begin{equation}
\bx_+(t)\equiv2\calA\bx(t)\equiv \bx(t)+\ri\calH\bx(t) \equiv 
      \label{trivariateanalytic}
    \re^ {\ri\phi (t)}\bQ_\bx(t) \begin{bmatrix}
        a(t) \\
        -\ri b(t)\\
        0
      \end{bmatrix}
\end{equation}
where $\calA$ is called the {\em analytic operator}.   This {\em defines} the six ellipse parameters on the right-hand side in terms of the three complex-valued quantities in $\bx_+(t)$; see \cite{lilly11-itsp} for explicit expressions.    Comparison with the kinematic model shows that (\ref{trivariateanalytic})  corresponds to setting $\varphi=0$ in (\ref{ellipsemodel}), thus $\bx(t)=\bx(t,0)$.    Thus when a signal $\bx(t)$ is observed, this is regarded as representing the position of one of a continuum of particles in an imaginary elliptical ring, the properties of which are assigned to the signal by the action of taking the analytic part.   We refer to this model for the evolution of $\bx(t)$ as the {\em modulated ellipse representation}.    

The ellipse parameters assigned to $\bx(t)$ by the analytic signal method are called the {\em canonical ellipse parameters}.   It is argued in 
 \citet{lilly10-itsp} and \citet{lilly11-itsp}, for the bivariate and trivariate cases respectively, that the ellipse parameters so determined represent the natural or intrinsic choice from the infinite family of possibilities.    The modulated ellipse representation explicitly captures joint variability by considering the components of $\bx(t)$ to be aspects of a single object.   This can  be seen as a natural time-varying generalization of the representation of the Fourier spectrum of a trivariate signal in terms of ellipse geometry as a function of frequency, as is common in optics \citep[e.g.][]{born}, oceanography \citep[e.g.][]{calman78a-jpo}, and seismology \citep[e.g.][]{park87b-jgr}, for example.  See \cite{lilly11-itsp} for further discussion of the ubiquity of the ellipse perspective in different fields.

\subsection{The analytic spectral matrix}\label{section:analytic}

The aggregate second-order structure of $\bx_+(t)$ is contained within its spectral matrix, or its Fourier transform, the autocovariance function. If $\bX(\omega)$ is the Fourier transform of $\bx(t)$, with $\bx(t) = \frac{1}{2\pi}\int_{-\infty}^\infty \bX(\omega)\re^{\ri\omega t} \, \rd \omega$, then we may define the one-sided deterministic spectral matrix as
\begin{equation}
\bS_{\bx}(\omega)\equiv 
\bX_+(\omega)\bX_+^H(\omega)
\end{equation}
where $\bX_+(\omega)\equiv 2U(\omega)\bX(\omega)$ is the Fourier transform of $\bx_+(t)$ with $U(\omega)$ being the unit step function.  Frequency-domain structure is then conveniently analysed in terms of the global moments of the spectral matrix
\begin{equation} \label{definemoments}
\overline \bS_n\equiv\frac{1}{2\pi}\int_{0}^\infty\omega^n  \bS_\bx(\omega) \,\rd\omega
\end{equation}
which are themselves matrix-valued.  The Fourier transform of the spectral matrix is the deterministic autocovariance function associated with the analytic signal $\bx_+(t)$
\begin{equation}\label{autocovdef}
\bR_\bx(\tau) \equiv  
\int_{-\infty}^\infty \bx_+(t+\tau) \bx_+^H(t) \rd t= \frac{1}{2\pi} \int_{0}^\infty  \bS_\bx(\omega) \re^{\ri \omega \tau } \rd\omega.
\end{equation}
An expansion of the autocorrelation matrix in terms of the time lag~$\tau$ then gives
\begin{equation}\label{rspair}
\bR_\bx(\tau)
 = \overline\bS_0 + \ri \tau \overline\bS_1 -\frac{1}{2}\tau^2 \overline\bS_2 + \cdots
\end{equation}
in which the frequency-domain moments defined in (\ref{definemoments}) appear as the coefficients.   Thus the global moment matrices $\overline \bS_n$ can equivalently be seen as describing  aspects of the aggregate frequency-domain structure of  $\bS_\bx(\omega)$, or else aspects of the aggregate time-domain structure captured in  $\bR_\bx(\tau)$ via powers of the time lag $\tau$.  For future reference, we introduce $S_\bx(\omega)\equiv \tr\,\bS_\bx(\omega)=\|\bX(\omega)\|^2$ as the trace of the spectral matrix.  This scalar-valued function of frequency, called the {\em joint analytic spectrum}, is the sum of the spectra of the analytic parts of all signal components.  The joint analytic spectrum $S_\bx(\omega)$ was examined by \citet{lilly10-itsp} and \citet{lilly11-itsp}, and this work will make an important reference point in what follows.

The use of the one-sided Fourier spectrum is standard in many fields, and it is important to emphasize that this involves the implicit use of the analytic signal.   Two crucial aspects are however not addressed by analysis of $\bS_\bx(\omega)$ or $\bR_\bx(\tau)$.  The first is that there is not yet a facility for investigating {\em time-varying} structure.  This is significant because many interesting signals exhibit properties that evolve with time, yet can have spectra which are indistinguishable from other, completely different signals.  The second is that is not obvious how to assign an {\em interpretation} to the moments $\overline\bS_n$.  One could discuss coherences, for example, but such an approach describes relationships between pairs of channels, rather than addressing the signal as a unified object with structure extending across all channels.    This suggests connecting the matrix-valued Fourier-domain moments $\overline\bS_n$ to the modulated ellipse model for time-varying structure, which is to be accomplished herein.

\subsection{The joint instantaneous moments}

The analytic signal is the basis for the creation of key time-varying quantities, called the {\em instantaneous moments}, that allow the moments of the spectrum of a univariate signal to be decomposed across time \citep[e.g.][]{gabor46-piee,ville48-cet,cohen89-ieee}.  In particular, the first three of these quantities---the {\em instantaneous amplitude,  frequency}, and {\em  bandwidth}---provide a powerful link between time variability and frequency-domain structure.   Here we follow \cite{lilly10-itsp} in
extending these quantities to describe the scalar-valued joint analytic spectrum $S_\bx(\omega)\equiv \tr\,\bS_\bx(\omega)$ of a multivariate signal.  Introduce
\begin{align}
\label{powermatrix}
\kappa_\bx^2(t) &\equiv \frac{1}{2}\|\bx_+(t)\|^2  
\\
\omega_\bx(t)&
\equiv \frac{\Imag\left\{\bx_+^H(t)\bx_+'(t)\right\}}{\|\bx_+(t)\|^2}  \\
\upsilon_\bx^2(t) \label{bandwidth}
&\equiv \frac{\left\|\bx_+'(t)\right\|^2}{\|\bx_+(t)\|^2}- \omega_\bx^2(t)
\end{align}
called the {\em joint instantaneous amplitude, frequency, and bandwidth}, respectively, of the multivariate signal $\bx(t)$.   These are found to satisfy the integral constraints
\begin{align}
\int_{-\infty}^{\infty}\kappa_\bx^2(t)\label{firstintegral}
\,\rd t  &=\frac{1}{2}\frac{1}{2\pi}\int_{0}^\infty S_\bx(\omega)\,\rd \omega\equiv\frac{1}{2}\,\calE_\bx\\
\frac{1}{\calE_\bx}\int_{-\infty}^{\infty}\omega_\bx(t)\|\bx_+(t)\|^2\,\rd t  &=\frac{1}{2\pi\calE_\bx}\int_{0}^\infty \omega S_\bx(\omega)\,\rd \omega\equiv\overline\omega_\bx
\end{align}
together with 
\begin{multline}\label{thirdintegral}
\frac{1}{\calE_\bx}\int_{-\infty}^{\infty} \left\{\upsilon_\bx^2(t)+\left[\omega_\bx(t)-\overline\omega_\bx\right]^2\right\} \|\bx_+(t)\|^2\,\rd t \\ =\frac{1}{2\pi\calE_\bx}\int_{0}^\infty \left[\omega-\overline\omega_\bx\right]^2 S_\bx(\omega)\, \rd \omega\equiv \overline\sigma_\bx^2
\end{multline}
and thus $\kappa_\bx(t)$, $\omega_\bx(t)$, and $\upsilon_\bx(t)$ respectively give the essential instantaneous or time-varying contributions to the zeroth-order, first-order, and second-order moments of the joint analytic spectrum  $S_\bx(\omega)$.  In the above, $\calE_\bx/2$ is the total energy of the signal $\bx(t)$, $\overline\omega_\bx$ is the mean frequency of  $S_\bx(\omega)$, and $\overline\sigma_\bx$ is the standard deviation about the mean frequency.  The name ``instantaneous bandwidth'' for $\upsilon_\bx(t)$ arises from the fact that $\overline\sigma_\bx$ is a measure of the signal's global Fourier bandwidth. 

It is shown in \cite{lilly10-itsp} that the joint instantaneous moments defined by (\ref{powermatrix})--(\ref{bandwidth}), reduce, for a univariate signal, to the standard definitions of the instantaneous amplitude, frequency, and bandwidth, respectively.  Therefore, as argued therein, the joint instantaneous moments generalize the univariate instantaneous moments to accommodate the moments of the joint analytic spectrum  $S_\bx(\omega)$ of a multivariate signal. The integral relations  (\ref{firstintegral})---(\ref{thirdintegral}) are precisely those which occur in the univariate case, see \S II~B of \citet{lilly10-itsp}.

As discussed in more detail later, the joint instantaneous moments  $\kappa_\bx(t)$, $\omega_\bx(t)$, and $\upsilon_\bx(t)$ can be shown to describe the variability of the oscillatory signal $\bx(t)$ in terms of the geometry of the modulated ellipse representation.  
This interpretation has been pursued by \citet{lilly10-itsp} and \citet{lilly11-itsp} for the bivariate and trivariate cases, respectively. However, in the trivariate case, the Hermitian instantaneous moment matrices  $\bS_n(t)$ each contain nine independent quantities, three real-valued quantities along the diagonal and three complex-valued quantities in the off-diagonal terms.  The joint instantaneous moments, being the traces of these matrices, describes only one of the nine components. In the following we will further generalize the instantaneous moments to matrix-valued quantities, and then obtain simple physical interpretations for the remaining eight terms.

\section{Unity of kinematic and physical moments}\label{section:unity}

Matrix-valued instantaneous spectral moments---kinematic quantities which generalize signal processing concepts of instantaneous amplitude,  frequency, and bandwidth to a multivariate signal---are introduced.  These are found to be identical to the hierarchy of {\em physical moments} of the evolving elliptical ring of particles which uniquely describes the observed signal, with several notable implications.

\subsection{Matrix-valued instantaneous moments}
  
The analytic signal method described in the previous section can be used as the basis for achieving a time-varying decomposition of the spectral matrix $\bS_\bx(\omega)$ of the signal $\bx(t)$,  building on the joint instantaneous moments of \citet{lilly10-itsp} that decompose the matrix trace. Define
\begin{eqnarray}
\label{powermatrixdef}
\bS_0(t)&\equiv&\bx_+(t)\,\bx_+^H(t)\\
\bS_1(t)&\equiv& 
\mherm\left\{-\ri\bx_+'(t)\,\bx_+^H(t)\right\}\\\label{bandwidthmatrixdef}
\bS_2(t)&\equiv&\bx_+'(t)\,{\bx_+'}\!\!^H(t) 
\end{eqnarray}
as the zeroth-order, first-order, and second-order {\em instantaneous moment matrices}, respectively, where $\mherm\, \bF \equiv (\bF+\bF^H)/2$ is the Hermitian part of the matrix~$\bF$.  Observe that these satisfy the integral constraint
\begin{equation} \label{integralconstraint}
\overline \bS_n\equiv\frac{1}{2\pi}\int_{0}^\infty\omega^n  \bS_\bx(\omega) \,\rd\omega = \int_{-\infty}^\infty\bS_n(t)\, \rd t
\end{equation}
and consequently the $\bS_n(t)$ give the instantaneous or local contributions that integrate to the corresponding global Fourier-domain moment matrices $\overline \bS_n$.   The $\bS_n(t)$ are not uniquely defined by the integral condition (\ref{integralconstraint}).  However it is sensible, and conventional in the analysis of univariate signals \citep[e.g.][]{cohen}, to constrain the instantaneous moments to have Hermitian symmetry, like the global moments to which they integrate; this leads to (\ref{powermatrixdef})--(\ref{bandwidthmatrixdef}) as the natural definitions.

The quantities $\bS_0(t)$, $\bS_1(t)$, and $\bS_2(t)$ provide normalized matrix-valued generalizations, appropriate for a multivariate signal, of  standard instantaneous moment quantities of a univariate signal---the instantaneous amplitude, frequency, and second central moment, respectively.  As these are all Hermitian matrices, each contains nine real-valued functions of time.   Taking the trace of (\ref{integralconstraint}), we find
\begin{equation}\label{recoverjointmoments}
\kappa_\bx^2(t) = \frac{1}{2}\tr\left\{\bS_0(t)\right\}, \quad 
\omega_\bx(t) = \frac{\tr\left\{\bS_1(t)\right\}}{\|\bx_+(t)\|^2}, \quad 
\upsilon_\bx^2(t) =\frac{\tr\left\{\bS_2(t)\right\}}{\|\bx_+(t)\|^2} -\omega_\bx^2(t). \end{equation}
and the traces of the matrix-valued instantaneous moments recover the scalar-valued joint instantaneous moments of  \citet{lilly10-itsp}.  The remaining terms in the $\bS_n(t)$ capture the time variability that accounts for the full second-order frequency-domain structure implicit in the spectral matrix $\bS_\bx(\omega)$.  Note that in defining the instantaneous moment matrices, it proves convenient to eschew the convention of normalizing by the signal power $\|\bx_+(t)\|^2$, as is done in the univariate case and for the joint instantaneous moments.  According to these definitions, the second instantaneous moment matrix $\bS_2(t)$ of $\bx(t)$ is also the zeroth instantaneous moment matrix of a different signal, the time derivative $\bx'(t)$ of the original signal.

\subsection{Physical moments of an elliptical ring of particles}\label{section:ellipse}

Using the analytic signal, an observed trivariate signal $\bx(t)$ can be described {\em as if} it were due to the trajectory traced out by a particle orbiting the periphery of a time-varying ellipse.   The time-varying position of any other particle within this hypothetical ellipse may then be found by a suitable choice of phase shift.  That is, we define a new signal $\bx(t,\varphi)$ as
\begin{equation}
\Real\left\{\re^{\ri\varphi} \bx_+(t)\right\} \label{trivariateanalyticfast}
 \equiv \bx(t,\varphi)
\end{equation}
and observe that this recovers the form of the kinematic model (\ref{ellipsemodel}).  By varying the phase shift $\varphi$, the trajectory of the observed particle $\bx(t)$---corresponding to $\bx(t,0)$---can be transformed into the trajectory $\bx(t,\varphi)$ of any other particle in the ellipse.   The modulated ellipse representation of a {\em signal} may therefore be viewed as specifying the time-varying position of an entire elliptical {\em ring} of particles, the {\em canonical ellipse}. This suggests computing the physical moments of the canonical ellipse, in order to understand how these are related to the kinematic moments.

Some physical moments of a ring of particles lying in a plane are defined in Table~\ref{table}.  The ring is taken to have mass density $\rho(\bx)=\rd M/\rd \ell$, with a total mass $M=\oint_C \rho(\bx)\,\rd\ell$ equal to unity, and with its centre of mass at the origin.  The moments are sorted in terms of their order---that is, the number of time derivatives involved---as well as whether they are scalar-valued, vector-valued, or tensor-valued.    There are two kinds of integrals in Table~\ref{table}, integrals in which the variable of integration is the scalar-valued differential arc length $\rd\ell$, occurring in the moments that lie along the main diagonal and counter diagonal of the table, and directed line integrals with a vector-valued variable of integration $\rd\bx$.  Moments of the latter type are defined to include a weighting by the linear mass density $\rho(\bx)$, while those of the former type are defined to include a factor of $\frac{1}{2\pi}$ but no mass weighting. It will emerge later that with the assumption of a unit-mass ring made here, both of these choices are equivalent; the distinction in  Table~\ref{table} is made in order that the moments take on familiar forms.

In Table~\ref{table} and henceforth, $\bI$ is the $3\times 3$ identity matrix.  Also we have introduced a notation to map a vector $\mathbf{f}\equiv \begin{bmatrix} f_x & f_y & f_z\end{bmatrix}^T$ into a skew-symmetric matrix,
\begin{equation}\label{crossdef}
\cross \, \mathbf{f}\equiv\begin{bmatrix} 0 & -f_z & f_y \\ f_z  & 0 & -f_x  \\ -f_y  & f_x  & 0\end{bmatrix}.
\end{equation}
The notation ``$\cross \,\mathbf{f}$'' is a reminder that we may write the cross product $\mathbf{f}\times \bg$ as the matrix multiplication $\left(\cross \,\mathbf{f}\right) \bg$, and thus $\cross \,\mathbf{f}$ may be referred to as the {\em cross-product matrix} associated with $\,\mathbf{f}$.  The cross-product matrix is in use in various communities \citep[e.g.][]{nour-omid91-cmame,liu08-ijiss,baksalary11-iam}, although a common notation does not seem to have yet emerged.

The categorization in Table~\ref{table} includes many familiar moments as well as some less familiar ones.  A measure of the typical distance from the ring to the origin is given by $\kappa_\bx(t)$.  The circulation $\Gamma_\bx(t)$ and kinetic energy $\frac{1}{2}\,\kappa_\bu^2(t)$ are familiar physical properties, as are the spatially averaged angular momentum vector $\angmom_\bx(t)$ and the moment of inertia tensor $\bm{\calI}_\bx(t)$.  The vector $\bn_\bx(t)$,  termed the {\em normal vector} following \cite{lilly11-itsp},  is more rarely encountered, although this integral appears in the calculation of the magnetic moment of a current loop \citep[e.g.][pp~64--65]{arfkenandweber}.  This vector points in the direction normal to the ring of particles, and as discussed in the preceding reference, from a variant of Stokes' theorem its magnitude is found to be $1/\pi$ times the area enclosed by any curve $C$.  At second order, $\bn_\bu(t)$ plays a similar role in describing the direction that is normal to the velocity, while $\bm{\calI}_\bu(t)$ is the kinetic energy tensor defined in such a way as to mimic the form of the moment of inertia tensor  $\bm{\calI}_\bx(t)$.  The most unusual quantity is the first-order tensor $\bm{\calJ}_\bx(t)$; this is a matrix-valued mixed product between the velocity and the position, complementing the scalar-valued mixed product, the circulation $\Gamma_\bx(t)$, and the vector-valued mixed product, the angular momentum $\angmom_\bx(t)$.

\begin{table}
\caption{Some moments of a ring of particles}
\begin{center}
\renewcommand{\arraystretch}{1.6}
\begin{tabular}{||l|lll||}\hline\hline
  &  Scalar-valued &  Vector-valued &   Matrix-valued \\\hline
 0 &   $\kappa^2_\bx(t) \equiv  \oint_C \|\bx\|^2 \rho\, \rd \ell$  &   $\bn_\bx(t) \equiv  \frac{1}{2\pi}\oint_C \bx \times \rd \bx$  & $\bm{\calI}_\bx(t)  \equiv  \oint_C\left(  \bI \|\bx\|^2-\bx\bx^T\right) \rho\, \rd \ell$ \\\hline
 1 & $\frac{\Gamma_\bx(t)}{2\pi} \equiv \frac{1}{2\pi} \oint_C \bu^T \rd \bx $  &    $\angmom_\bx(t)\equiv  \oint_C\bx\times\bu\, \rho\, \rd \ell$ &  $\bm{\calJ}_\bx(t)  \equiv \mherm \frac{1}{2\pi}\oint_C (\cross\, \bu)^T \cross\, \rd \bx$ \\\hline
 2 &   $\kappa_\bu^2(t) \equiv   \oint_C\|\bu\|^2\rho\, \rd \ell$  &  $\bn_\bu(t) \equiv  \frac{1}{2\pi}\oint_C \bu \times \rd \bu$  &   $\bm{\calI}_\bu(t)  \equiv  \oint_C\left( \bI \|\bu\|^2-\bu\bu^T\right) \rho\, \rd \ell$\\
\hline\hline
\end{tabular}\end{center}\longcaption{Moments of a time-varying closed curve of particles $C(t)$ having unit mass, with  $\rho=\rho(\bx)$ being the linear mass density along the curve. The particle position is $\bx$ and the velocity is $\bu=\bu(\bx)$. Directed line integrals with a vector-valued variable of integration are taken by convention in the right-hand sense. The moment order is shown at the left.}
\label{table}
\end{table}

The main result of this work is to show that the first three instantaneous moments matrices, defined in (\ref{powermatrixdef})--(\ref{bandwidthmatrixdef}) in order to identify the instantaneous contributions to the Fourier-domain moments of the spectral matrix, take the forms
\begin{eqnarray}\label{S0expand}
\frac{1}{2}\,\bS_0(t)&=&\kappa_\bx^2(t)\bI  - \ri \frac{1}{2}\,\cross\,\bn_\bx(t) -\bm{\calI}_\bx(t) \\\label{S1expand}
\frac{1}{2}\,\bS_1(t)&=&\frac{\Gamma_\bx(t)}{2\pi}\,\bI+\ri \frac{1}{2}\,\cross\,\angmom_\bx(t)-\bm{\calJ}_{\!\bx}(t)\\
\frac{1}{2}\,\bS_2(t)&=&\kappa_\bu^2(t)\bI - \ri\frac{1}{2} \,\cross\,\bn_\bu(t)-\bm{\calI}_\bu(t) \label{S2expand}
\end{eqnarray}
and consequently can be interpreted as containing the physical moments of the canonical ellipse.  As will be shown later, each of these complex-valued Hermitian matrices contains, from left to right, a scalar-valued portion, a vector-valued portion, and a tensor-valued portion, respectively.  The instantaneous spectral moments, and the physical moments of the canonical ellipse, are therefore identical.   It is not at all obvious that this should be the case.  This result is termed the {\em unity of kinematic and physical moments}.

Some implications of this result are pursued in the remainder of this section.  The proof of the unity of moments is postponed until \S \ref{section:proof}.   Detailed interpretation of individual terms in the instantaneous moments will be addressed in \S \ref{section:moments}.

\subsection{Interpretation of instantaneous contributions to the spectral matrix}

The unity of moments provides a rich set of geometric tools for analyzing the time-varying structure of a bivariate or trivariate signal.   The structure implicit in the matrix-valued instantaneous moments can now be described using a set of physical concepts, such as moment of inertia and angular momentum, with which we are already intimately familiar.   This interpretation provides us with new insight into the spectral matrix itself.   

For example, the zeroth-order global moment  $\overline{\bS}_0$ of the spectral matrix can be decomposed as
\begin{multline}
\frac{1}{2}\,\overline\bS_0 =\frac{1}{2}\int_{-\infty}^\infty \bS_\bx(t) \, \rd t
 =\overline{\kappa^2_\bx} \,\bI - \ri \frac{1}{2}\,\,\cross\,\overline{\bn}_\bx-\overline{\bm{\calI}}_\bx \\
 =\bI \int_{-\infty}^\infty \kappa_\bx^2(t)\, \rd t  -\ri  \frac{1}{2}\,\int_{-\infty}^\infty \cross\,\bn_\bx(t)\, \rd t- \int_{-\infty}^\infty \bm{\calI}_\bx(t) \, \rd t
\end{multline}
where the averages on the first line are defined as the corresponding integrals over instantaneous quantities on the second line.   The scalar, vector, and tensor portions of the zeroth-order instantaneous moment matrix are therefore seen as being respectively built up from instantaneous contributions from the squared instantaneous amplitude, the normal vector, and the moment of inertia tensor of the canonical ellipse.   The variability of $\bS_0(t)$ therefore captures, in its different structure components, aspects of the geometric variability of the canonical ellipse which imprint themselves directly onto the spectrum.  This lends itself to a wide variety of analysis applications.  One could, for example, temporally average $\bS_0(t)$ and the other instantaneous moment matrices during different time periods in order to quantity the contributions to the spectral matrix due to distinct modes of observed behaviour.

More generally we may see the scalar-valued, vector-valued, and tensor-valued portions of the instantaneous moment matrices can be seen as decomposing the corresponding portions of the spectral matrix itself, which can be written
\begin{multline}\label{Sexpand}
\bS_\bx(\omega)=\bX_+(\omega)\bX_+^H(\omega)   =  S_\bx(\omega)\bI- \ri \frac{1}{2}\cross\Imag\left\{\bX_+(\omega)\times\bX_+^*(\omega)\right\}\\
-\mathrm{herm}\Real\left\{ \left[\cross \bX_+(\omega)\right]^H\cross\bX_+(\omega) \right\}
\end{multline}
as will be shown in the next section.   Observe that the quadrature spectrum---the imaginary part of $\bS_\bx(\omega)$---is entirely wrapped up in the vector-valued term in (\ref{Sexpand}).   Thus, from (\ref{S0expand})--(\ref{S2expand}), we see that the zeroth-order, first-order, and second-order instantaneous moments associated with the quadrature spectrum are the  cross-product matrices of the normal vector of the canonical ellipse $\bn_\bx(t)$, the negative of the angular momentum vector $\angmom_\bx(t)$, and the normal vector of the time derivative of the canonical ellipse $\bn_\bu(t)$, respectively.  This gives a new way of interpreting the information contained within the quadrature spectrum.   Alternatively, the components of instantaneous moment matrices may be viewed as providing different contributions to the autocorrelation matrix $\bR_\bx(\tau)$ associated with various powers of time lag $\tau$, which we see from inserting  (\ref{S0expand})--(\ref{S2expand}) into (\ref{rspair}).


\subsection{Recovery of physical ellipse parameters}

A second important implication concerns those signals for which the physical moments in Table~\ref{table} are in fact themselves the sought-after quantities.   Imagine that there exists a physical  ring of particles in space,  evolving according to the kinematic ellipse model (\ref{ellipsemodel})  for some choice of time-varying ellipse parameters.   A broad class of physical systems could be described in this manner, including  observations of  oceanic vortices using freely drifting or ``Lagrangian'' instruments \citep{lilly06-npg,lilly12-itsp}.    In general, the trajectory of a particle in such a system will  become associated,  after we take the analytic part,  with a set of canonical ellipse parameters  that are {\em not} the same  as the physical system that generated the signal.   If we put an arbitrarily chosen set of time-varying ellipse parameters into the  modulated ellipse representation (\ref{trivariateanalytic}),  there is no reason to assume that this will generate an analytic signal, and yet we know that the canonical ellipse parameters will do so by definition.     If on the other hand the physical ellipse parameters are such that they {\em do} generate an analytic signal via (\ref{trivariateanalytic}),  then the analytic signal method will exactly recover them from observed signal.

This means that in some circumstances,  the trajectory of a single  {\em particle} is sufficient to recover exactly the time-varying properties of the entire {\em  system}, and these {\em physical moments} integrated across time determine the structure of the observed spectral matrix.  Denote the physical ellipse parameters as $\widetilde a(t)$,   $\widetilde b(t)$,  etc., in order to emphasize that these are prescribed and not necessarily the canonical ellipse parameters of the generated signal.   It follows immediately from  the work of  \citet{bedrosian63-ire} for univariate signals---the so-called ``Bedrosian's theorem''---that a sufficient condition for  the physical ellipse parameters to be canonical is that all components of the $3 \times 2$ matrix
\begin{equation}
\widetilde \bQ_\bx(t) \begin{bmatrix} \,\widetilde a(t) & 0 \\
 0 & \widetilde b(t)  \\ 0 & 0
 \end{bmatrix}\equiv \bJ_3(\widetilde \alpha(t))\, \bJ_1(\widetilde \beta(t))\, \bJ_3(\widetilde\theta(t)) \begin{bmatrix} \,\widetilde a(t) & 0 \\
 0 & \widetilde b(t)  \\ 0 & 0
 \end{bmatrix}
\end{equation}
are supported entirely on lower (i.e., less positive)  frequencies than is the phase function $e^{\ri \widetilde \phi(t)}$.   This amounts to a statement that the ellipse geometry evolves more slowly than the circulation of the particles around the ellipse.   

Under these conditions, Bedrosian's theorem implies that the analytic operator ``passes through'' the matrix, giving
\begin{equation}
\bx_+(t)= 2 \calA \left\{
 \widetilde \bQ_\bx(t)
        \begin{bmatrix} \widetilde a(t) & 0 \\
 0 & \widetilde  b(t)  \\ 0 & 0
 \end{bmatrix}\begin{bmatrix} \cos  \widetilde \phi(t) 
 \\ \sin \widetilde  \phi(t) 
   \end{bmatrix}  \right\}
=\re^ {\ri\widetilde \phi (t)}\widetilde \bQ_\bx(t) \begin{bmatrix}
        \widetilde a(t) \\
        -\ri \widetilde  b(t)\\
        0
      \end{bmatrix}
\end{equation}
and since this matches the form of (\ref{trivariateanalytic}), we see that the prescribed and canonical ellipse parameters are identical.  As  it is reasonable to believe that a subset of oceanic vortices may satisfy or approximately satisfy this condition of slow variation, it appears  that the link between the observed trajectory of a single Lagrangian particle,   and physical structures in the vicinity of this observation point, is considerably more direct  than has been previously recognized.  Further investigation of this point,  and  comparison between the domains  over which the ellipse parameters  are canonical and the domains over which dynamical solutions exist,  is called for but its outside the scope of this work.

\section{Proof of the unity of moments}\label{section:proof}

The proof of the unity of moments consists of three components: a Hermitian matrix expansion; a relationship between products of analytic signals and a phase averaging operator; and a reduction of integrals around the ellipse to a phase average.  

\subsection{A matrix expansion using complex vectors}

In what follows we will make use {\em complex-valued} three-vectors, which warrants some discussion as these objects are somewhat rarely encountered.  However,  their use dates back over a century to Gibbs and the beginning of modern vector analysis \citep[see ``Note on bivector analysis'', p. 84---90 of][]{gibbs06-vol2}. Necessary basic facts are given here; further details may be found in \citet{gibbs06-vol2}, as well as \citet{lindell}  and \citet{bivectors}.  The cross-product of two real-valued or complex-valued vectors  $\mathbf{f}\equiv \begin{bmatrix} f_x & f_y & f_z\end{bmatrix}^T$  and  $\bg\equiv \begin{bmatrix} g_x & g_y & g_z\end{bmatrix}^T$  is defined as
\begin{equation}
\mathbf{f}\times\bg\equiv
\left(f_y g_z-f_z g_y  \right)\mathbf{i}
-\left(f_x g_z-f_z g_x  \right)\bj
+\left(f_x g_y-f_y g_x  \right)\bk
\label{crossproductdef}
\end{equation}
where $\mathbf{i}$, $\bj$, and $\bk$ are the unit vectors along the $x$, $y$, and $z$-axes, respectively.  Observe that there is no conjugation in this definition.  The cross product of a complex vector with itself vanishes, $\mathbf{f}\times\mathbf{f}=\bzero$, but the cross product of a vector with its own complex conjugate has a nonzero purely imaginary part, $\mathbf{f}\times\mathbf{f}^* = 2 \ri \Imag\mathbf{f}  \times \Real \mathbf{f}$.

The following identities involving the cross-product matrix of complex-valued vectors may be readily verified:
\begin{align}
\left(\cross \,\mathbf{f}\right)^H\!\cross\,\bg&=\left(\mathbf{f}^H\bg\right)\!\bI-\bg\,\!\mathbf{f}^H \label{crossofcross}\\
\frac{1}{2}\,\cross\Imag\left\{\mathbf{f}\times  \mathbf{g}^*\right\}&=-\Imag\left\{\mherm\,\bg\,\!\mathbf{f}^H\right\}
\label{crossskew}
\end{align}
where the cross-product matrix $\cross \,\mathbf{f}$ of a real-valued or complex-valued vector $\mathbf{f}$ is defined by (\ref{crossdef}).   Other identities involving the cross-product matrix may be found in  \citet{liu08-ijiss} and \citet{baksalary11-iam}. Combining  (\ref{crossofcross}) and (\ref{crossskew}) yields an expansion for a Hermitian matrix, which we write without loss of generality as $\mherm\,\bg\,\!\mathbf{f}^H$,  as
\begin{equation}\label{hermexpand}
\mherm\, \bg\,\!\mathbf{f}^H =\overset{(\ri)}{\overbrace{\Real\left\{\mathbf{f}^H\bg\right\}}}\,\bI
-\overset{(\ri\ri)}{\overbrace{\ri\frac{1}{2}\,\cross\Imag\left\{\mathbf{f}\times  \mathbf{g}^*\right\}}}
-\overset{(\ri\ri\ri)}{\overbrace{\mherm\Real\left\{\left(\cross \,\mathbf{f}\right)^H\!\cross\,\bg\right\}}}.
\end{equation}
This consists of a scalar-valued product (i), a vector-valued product (ii),  and a matrix-valued product (iii)  of  $\mathbf{f}$ and $\bg$.  The scalar-valued and matrix-valued portions (i) and (iii) are real-valued, while the vector-valued portion (ii) is imaginary.  Note that the traces of the three quantities on the right-hand side are $3\Real\left\{\mathbf{f}^H\bg\right\}$, zero, and $-2\Real\left\{\mathbf{f}^H\bg\right\}$, respectively; thus the coefficient of $\bI$ in (\ref{hermexpand}), which we refer to as the scalar-valued portion of the matrix, is also the matrix trace.  

Using (\ref{hermexpand}), we now expand the instantaneous spectral matrices as
\begin{multline}\label{hermexpandS0}
\bS_0(t)=\|\bx_+(t)\|^2\,\bI
-\ri\frac{1}{2}\,\cross\Imag\left\{\bx_+(t)\times  \bx_+^*(t)\right\}\\-\mherm\Real\left\{\left[\cross \,\bx_+(t)\right]^H\!\cross\,\bx_+(t)\right\}
\end{multline}
for the zeroth moment matrix and 
\begin{multline}\label{hermexpandS1}
\bS_1(t)=\Imag\left\{\bx_+^H(t)\bx_+'(t) \right\}\bI
+\ri\frac{1}{2}\,\cross\Real\left\{\bx_+(t)\times  {\bx_+'}\!\!^*(t)\right\}\\-\mherm\Imag\left\{\left[\cross \,\bx_+(t)\right]^H\!\cross\, \bx_+'(t)\right\}
\end{multline}
for the first moment matrix, and so forth.  In the following subsections we will see how to relate these various terms to the physical moments.

\subsection{Phase averaging}\label{phaseaveraging}

Products of the components of the analytic signal $\bx_+(t)$ or its derivatives implicitly evaluate integrals around the periphery of the canonical ellipse, as will now be shown.  Let $f(t)$ be the product of some number, say $M$, components of $\bx(t)$ or one of its derivatives of some order 
\begin{equation} \label{fform}
f(t)= x_{i_1}^{(n_1)}(t)\, x_{i_2}^{(n_2)}(t)\cdots x_{i_M}^{(n_M)}(t)
\end{equation}
where the superscript ``${(n_m)}$'' indicates the $n_m$th-order derivative with respect to time; here we assume derivatives of up to the required order exist and are finite.  The \emph{phase average} of $f(t)$, defined as 
\begin{equation}\label{phaseaverage}
\overline{f(t)}^{\,\varphi}\equiv
\frac{1}{2\pi}\int_{-\pi}^{\pi}x_{i_1}^{(n_1)}(t,\varphi)\, x_{i_2}^{(n_2)}(t,\varphi)\cdots x_{i_M}^{(n_M)}(t,\varphi)\,\rd\varphi
\end{equation}
produces a smoothed, but still time-varying, version of some property $f(t)$ by averaging over all particles in the ellipse.  The terms appearing in the integrand are components of the phase-shifted version of the signal $\bx(t,\varphi)$ defined in (\ref{trivariateanalyticfast}) or one of its partial derivatives with respect to time, which we indicate with the superscripts.   Effectively, (\ref{phaseaverage}) allows us to separates variation {\em around} the ellipse from changes in the ellipse geometry; variations on this idea have been used previously by \cite{lilly06-npg} and \cite{schreier08b-itsp}.

For a given function $f(t)$ of the form (\ref{fform}), the phase average is constructed by algebraically calculating expressions for the $M$ terms in $f(t)$ using the modulated ellipse representation of the phase-shifted signal (\ref{trivariateanalyticfast}), and then performing the indicated average in (\ref{phaseaverage}).    As an example of a quadratic product ($M=2$), the phase-averaged squared magnitude of the real-valued vector $\bx(t)$ is given by
\begin{multline}\label{phaseavgexample}
\overline{\|\bx(t)\|^2}^{\,\varphi}=\overline{|x(t)|^2}^{\,\varphi}+\overline{|y(t)|^2}^{\,\varphi}+\overline{|z(t)|^2}^{\,\varphi}
=\frac{1}{2\pi}\int_{-\pi}^{\pi}\!\!
\bx^T(t,\varphi)\bx(t,\varphi) \rd\varphi\\
=\frac{1}{2\pi}\int_{-\pi}^{\pi}\!\!
\left[a^2(t)\cos^2\left(\phi(t)+\varphi\right)+
b^2(t)\sin^2\left(\phi(t)+\varphi\right)\right]\rd\varphi\\
= \frac{a^2(t)+b^2(t)}{2} = \frac{1}{2}\|\bx_+(t)\|^2   =\kappa_\bx^2(t)
\end{multline}
which is one-half the instantaneous power of the analytic signal vector. This does not depend on the orbital phase $\phi(t)$, while by contrast the unaveraged quantity
\begin{equation}
\|\bx(t)\|^2= a^2(t)\cos^2(\phi(t))+b^2(t)\sin^2(\phi(t))
\end{equation}
varies rapidly as the particle orbits the ellipse, even for fixed ellipse geometry.

In this way the analytic signal can be used to greatly simplify calculations involving phase averages of the real-valued signal.  Let $\xi_1(t)$ and $\xi_2(t)$ each be an element of $\bx_+(t)$ or one of its derivatives, so that $\Real\xi_1(t)$ and $\Real\xi_2(t)$ each represent one of the  $M$ terms in  (\ref{fform}).   $\xi_1(t)$ and $\xi_2(t)$ are analytic because taking the analytic part commutes with differentiation.   The phase averages of the products of real or imaginary parts of $\xi_1(t)$ and $\xi_2(t)$ are found to be
\begin{align}\label{realimagtheorem}
\overline{\Real\xi_1(t)\Real\xi_2(t)}^{\,\varphi}=\overline{\Imag\xi_1(t)\Imag\xi_2(t)}^{\,\varphi}&=\frac{1}{2}\,\Real\left\{\xi_1(t)\,\xi_2^*(t)\right\}\\
\overline{\Imag\xi_1(t)\Real\xi_2(t)}^{\,\varphi}=-\overline{\Real\xi_1(t)\Imag\xi_2(t)}^{\,\varphi}&=
\frac{1}{2}\,\Imag\left\{\xi_1(t)\,\xi_2^*(t)\right\}\label{realimagtheorem2}
\end{align}
and therefore, significantly, these phase averages can be calculated by simply taking the real or imaginary part of the Hermitian product of the analytic functions.    In the above example, direct calculation in (\ref{phaseavgexample}) shows that $\overline{\|\bx(t)\|^2}^{\,\varphi}= \|\bx_+(t)\|^2/2$.  However, this follows at once from (\ref{realimagtheorem}), because choosing $\xi_1(t)=\xi_2(t)=x_+(t)$ we have $\overline{|x(t)|^2}^{\,\varphi}=|x_+(t)|^2/2$ and similarly for $y(t)$ and $z(t)$.  

To verify the quadratic phase averaging results (\ref{realimagtheorem}) and (\ref{realimagtheorem2}), write the functions $\xi_1(t)$ and $\xi_2(t)$ in the form
\begin{equation}
\xi_1(t) = \re ^{\ri\phi(t)} \left[c_1(t)+\ri d_1(t)\right],\quad\quad
\xi_2(t) = \re ^{\ri\phi(t)} \left[c_2(t)+\ri d_2(t)\right]
\end{equation}
where $c_1(t)$, $d_1(t)$, $c_2(t)$, and $d_2(t)$  are all-real valued and do not depend explicitly on the phase $\phi(t)$, although they may depend upon derivatives of $\phi(t)$.    To see  (\ref{realimagtheorem}), note that the product of the real parts is given by
\begin{multline}
\Real\xi_1(t)\Real\xi_2(t)=
\frac{1}{2}\left[c_1(t)c_2(t)+d_1(t)d_2(t)\right] \\+
\frac{1}{2}\left[c_1(t)c_2(t)-d_1(t)d_2(t)\right]\cos\left( 2\phi(t) \right)\\
-\frac{1}{2}\left[c_1(t)d_2(t)+c_2(t)d_1(t)\right]\sin\left( 2\phi(t)\right)\label{abouttoaverage}
\end{multline}
where we have made use of a trigonometric identity.   At the same time, we have $\Real\left\{\xi_1(t)\xi_2^*(t)\right\}=c_1(t)c_2(t)+d_1(t)d_2(t)$, and thus carrying out the phase average in (\ref{abouttoaverage}) obtains (\ref{realimagtheorem}).     Similarly $\Imag\xi_1(t)\Imag\xi_2(t)$ is found to contain a constant term, equal to the constant term in  (\ref{abouttoaverage}), plus oscillatory terms.  We then find
\begin{multline}\label{imagtheorem}
\overline{\Imag\xi_1(t)\Real\xi_2(t)}^{\,\varphi}=-\overline{\Real\xi_1(t)\Imag\xi_2(t)}^{\,\varphi}= \frac{1}{2}\left[c_2(t)d_1(t)-c_1(t)d_2(t)\right]\\=
\frac{1}{2}\,\Imag\left\{\xi_1(t)\,\xi_2^*(t)\right\}
\end{multline}
for the phase average of the product of real and imaginary parts, verifying (\ref{realimagtheorem2}).    In \ref{appendix:phaseaveraging}, a related phase averaging result for a quartic ($M=4$) moment is derived for future reference.

\subsection{Phase averages with scalar-valued variables of integration}

The physical moments in Table~\ref{table} can be related to phase averages.   The five integrals with the scalar-valued variable of integration $\rd\ell$ will be considered first.  With a change of variables, the moment of inertia $\bm{\calI}_\bx(t)$ defined in  Table~\ref{table} becomes
\begin{equation}\label{willbeaverage}
\bm{\calI}_\bx(t) = \int_{-\pi}^\pi \left[\,\bI\|\bx(t,\varphi)\|^2 - \bx(t,\varphi)\bx^T(t,\varphi)\right]
\frac{\rd M}{\rd \varphi}
\,\rd\varphi.
\end{equation}
As all particles have the same mass by assumption, the differential mass per differential particle $\rd M / \rd \varphi$ is a constant. Since we have set the ring to be unit mass, 
\begin{equation}
M=\oint_C \rho(\bx)\, \rd\ell=\int_{-\pi}^\pi \frac{\rd M}{\rd \ell} \frac{\rd \ell}{\rd \varphi} \,\rd\varphi = \int_{-\pi}^\pi \frac{\rd M}{\rd \varphi} \,\rd\varphi = 2\pi \frac{\rd M}{ \rd \varphi}=1
\end{equation}
and so $\rd M / \rd \varphi=1/(2\pi)$.  Consequently (\ref{willbeaverage}) becomes a phase average,  leading to 
\begin{align}
\bm{\calI}_\bx(t)&=  \overline{\,\bI \|\bx(t)\|^2-\bx(t)\bx^T(t)}^{\,\varphi} \\
&=\frac{1}{2}\,\bI\left\|\bx_+ (t)\right\|^2 -\frac{1}{2}\Real\left\{\bx_+(t)\bx_+^H(t)\right\}
=\frac{1}{2}\Real\left\{\left[\cross\,\bx_+(t)\right]^H\cross\,\bx_+(t)\right\}\label{canonicalI}
\end{align}
with the second line following from applying the phase-averaging result (\ref{realimagtheorem}) separately to each component of the matrix, and using (\ref{crossofcross}) to obtain the last equality.   Substituting for $\bx_+(t)$ from (\ref{trivariateanalytic}) into (\ref{canonicalI}), we find the moment of inertia tensor obtains the familiar form
\begin{equation}
\bm{\calI}_\bx(t)=\frac{1}{2}\,\bQ_\bx(t)\begin{bmatrix} b^2(t) & 0  & 0 \\ 0 & a^2(t) & 0\\
0 & 0 & a^2(t)+b^2(t) \end{bmatrix}\bQ_\bx^T(t).
\end{equation}
It is worth pointing out that the assumption of constant mass density as a function of particle label, i.e. constant $\rd M / \rd \varphi$, implies that the linear mass density $\rho(\bx)=\rd M /\rd \ell$ of the canonical ellipse is not constant; the latter changes as particles speed up or slow down in their flow along the ellipse periphery. 

In the same way, integrating the angular momentum over the ellipse as defined in  Table~\ref{table} leads~to
\begin{equation}
\angmom_\bx(t)=\label{angmomdef}
\overline{\bx (t) \times \bx'(t)}^{\,\varphi}=\frac{1}{2}\Real\left\{\bx_+(t) \times{\bx_+'}\!\!^*(t)\right\}
\end{equation}
after reducing the integral to a phase average, then applying (\ref{realimagtheorem}) separately to each component of the vector to obtain the final expression.   For the scalar-valued zeroth moment $\kappa_\bx^2(t)$ we may write 
\begin{equation}
\kappa_\bx^2(t) = \oint_C \|\bx\|^2 \rho(\bx)\, \rd \ell = \overline{\left\|\bx(t)\right\|^2}^\phi
=\overline{\left\|\bx(t)-\overline{\bx(t)}^\phi\right\|^2}^\phi
\end{equation}
on account of the fact that the phase average of $\bx(t)$ vanishes.  In this form we see that $\kappa_\bx^2(t)$ is the {\em variance} of the position vector $\bx(t)$ around the ellipse periphery.   As $\kappa_\bu^2(t)$ and  $\bm{\calI}_\bu(t)$ are identical in form to $\kappa_\bx^2(t)$ and  $\bm{\calI}_\bx(t)$ respectively, this verifies five of the nine correspondences in the spectral matrix expansions (\ref{S0expand})--(\ref{S2expand}).

\subsection{Phase averages with vector-valued variables of integration}

To verify the directed line integrals in (\ref{S0expand})--(\ref{S2expand}), we first make an observation regarding the interpretation of the Hilbert transform.  The obvious interpretation of the Hilbert transform is that it provides the particle position for a partner or ``ghost'' particle that lags ninety degrees behind the observed particle, since we have
\begin{equation}\label{shiftedhilbert}
\calH\bx(t)=  \Real\left\{-\ri \bx_+(t) \right\} = \Real\left\{  \re^ {\ri\phi (t)-\ri\pi/2}\,\bQ_\bx(t) \left[\begin{array}{c}
        a(t)  \\
        -\ri b(t)\\
        0
      \end{array}\right]\right\}.
\end{equation}
This means that the Hilbert transform contains (or more accurately, assigns) nonlocal information about what is happening at another point in the ellipse. However, taking the partial derivative of (\ref{ellipsemodel}) with respect to phase $\varphi$, 
one finds
\begin{equation}\label{hilbertphase}
\frac{\partial}{\partial\varphi}\,\bx(t,\varphi)  =  \bQ_\bx(t)\begin{bmatrix}   -a(t) \sin (\phi(t)+\varphi) \\ b(t) \cos (\phi(t)+\varphi) \end{bmatrix}=- \calH \bx(t,\varphi)
\end{equation}
which shows that the rate of change of the position vector with variations in phase is the same as negative of the Hilbert transform. (In the expression $\calH \bx(t,\varphi)$, it is understood that $\calH$ acts on the time variable.) This means that the Hilbert transform also contains (or assigns) local information, expressing the variation in the phase-shifted signal vector $\bx(t,\varphi)$ as we move from one particle to the next.

This observation allows us to simplify the computation of directed line integrals.  For example, the circulation of the canonical ellipse becomes
\begin{equation}\label{definegamma}
\Gamma_{\!\bx}(t)\equiv  \oint_{C} \bu^T \rd\bx = 
\int_{-\pi}^\pi\left[\frac{\partial}{\partial t}\bx(t,\varphi)\right]^T\frac{\partial\bx}{\partial\varphi}(t,\varphi) \,\rd \varphi
\end{equation}
after a change of variables.  Substituting  for $\frac{\partial}{\partial\varphi}\,\bx(t,\varphi)$ from  (\ref{hilbertphase}) then leads to a phase average, and we obtain
\begin{equation}
\Gamma_{\!\bx}(t) =  -2 \pi\, \overline{\,[\bx'(t)]^T\calH \bx(t)}^{\,\varphi} = \pi \Imag \left\{\bx_+^H(t)\bx_+'(t) \right\} =\pi \,\tr\,\bS_1(t)
\end{equation}
after making use of (\ref{realimagtheorem2}).  This verifies the trace of  $\bS_1(t)$ in (\ref{S1expand}).  Similarly the integral for the normal vector $\bn_\bx(t)$  becomes, again using  (\ref{hilbertphase})  and  (\ref{realimagtheorem2}), 
\begin{multline}\label{nexpression}
\bn_\bx(t)= \frac{1}{2\pi}\oint_C \bx\times \rd \bx 
=\frac{1}{2\pi}\int_{-\pi}^\pi\bx(t,\varphi)\times \frac{\partial}{\partial\varphi}\,\bx(t,\varphi) \,\rd \varphi \\=
-\overline{\, \bx(t) \times \calH \bx(t)}^{\,\varphi} = \overline{\,  \calH \bx(t)\times \bx(t) }^{\,\varphi} =\frac{1}{2} \Imag\left\{\bx_+(t)\times\bx_+^*(t)\right\}
\end{multline}
which verifies the imaginary part of $\bS_0(t)$ in (\ref{S0expand}).   In the same way we find
\begin{align}
\bn_\bu(t)&= \overline{\,  \calH \bx'(t)\times \bx'(t) }^{\,\varphi} =\frac{1}{2} \Imag\left\{\bx_+'(t)\times{\bx_+'}\!\!^*(t)\right\}
\\ \label{Jdefmatrix}
 \bm{\calJ}_{\!\bx}(t)&=\mherm \,\overline{\left[\cross\,\bx'(t)\right]^T\!\cross\,\calH\bx(t)}^{\,\varphi} =
\frac{1}{2}\, \mherm\,\Imag \left\{\left[\cross\,\bx_+(t)\right]^H \cross\,\bx_+'(t)\right\}
\end{align}
thus verifying the four correspondences in (\ref{S0expand})--(\ref{S2expand}) involving directed line integrals.  This completes the proof of the unity of moments.

\section{Interpretation of instantaneous moments}\label{section:moments}

The recognition that the instantaneous spectral moments are identical to physical moments of an elliptical ring of particles calls for a close examination of key physical quantities---in particular, the circulation, angular momentum, and kinetic energy---and their relationships to instantaneous amplitude, frequency, and bandwidth.

\subsection{Instantaneous frequency and circulation}

The most striking result from the unity of moments concerns the trace of the first-order matrix  $\bS_1(t)$.  Comparing  (\ref{recoverjointmoments}) with (\ref{S1expand}) we find
\begin{equation}\label{frequencyandcirc}
\frac{1}{2\pi} \,\Gamma_{\!\bx}(t)= \kappa_\bx^2(t) \omega_\bx(t) = \frac{1}{2} \Imag\left\{ \bx_+^H(t) \bx_+'(t)\right\}
\end{equation}
which shows that the circulation of the canonical ellipse is identical to the joint instantaneous frequency weighted by the squared signal amplitude.   This is significant because it unifies two key quantities, the instantaneous frequency and the circulation, that are in use by two largely disparate communities.   On the one hand, it provides new insight into what the instantaneous frequency \emph{is}; on the other, it indicates  how the time-varying circulation may be directly estimated by the analytic signal method.  It follows that the global mean frequency is
given by
\begin{equation}
\overline \omega_\bx \equiv\frac{1}{2\pi\calE_\bx}\int_{0}^\infty\omega \, \tr\,\bS_\bx(\omega) \,\rd\omega = \frac{1}{\calE_\bx}\int_{-\infty}^\infty\frac{1}{\pi}\,\Gamma_{\!\bx}(t)\, \rd t
\end{equation}
and therefore is interpreted as $1/\pi$ times the time-integrated circulation of the canonical ellipse. 

It is relevant to mention that for the canonical ellipse, $\overline\omega_\bx$, and thus the time-averaged circulation, must always be positive.  While circulation can be either positive or negative in general, here we have made an implicit coordinate system choice regarding the sign of the vertical axis with respect to the plane containing the ellipse.  A negative circulation can be transformed into a positive circulation through a three-dimensional rotation that changes the sign of the vertical axis. It is possible, however, that $\omega_\bx(t)$ or $\Gamma_\bx(t)$ may {\em locally} reverse signs and become negative.

If we imagine that there is a two-dimensional flow in the entire area bounded by the canonical ellipse, then by Stokes' theorem, the canonical circulation $\Gamma_{\!\bx}(t)$ is equivalent to the vorticity integrated over this elliptical area.  The spatially-averaged vorticity $\zeta_\bx(t)$ is given by
\begin{equation}\label{vorticity}
\zeta_\bx(t)\equiv
\frac{\Gamma_{\!\bx}(t)}{\pi a(t) b(t)} = \omega_\bx(t)\,\frac{a^2(t)+b^2(t)}{a(t) b(t)} = \frac{2\omega_\bx(t)}{\sqrt{1-\lambda^2(t)}}
\end{equation}
where $\lambda(t) \equiv  \frac{a^2(t)-b^2(t)}{a^2(t)+b^2(t)}$ is a measure of the ellipse shape, which, like the eccentricity, varies between zero for circular motion and unity for linear motion.  Observe that for a circular vortex, $\lambda(t)=0$, and (\ref{vorticity}) reduces to $\zeta_\bx(t)=2\omega_\bx(t)$.  This echoes an elementary result that the vorticity of a vortex in solid-body rotation is twice its angular frequency;  thus (\ref{vorticity}) may be seen as a generalization of this result to a link between the time-varying canonical ellipse and its instantaneous frequency.

A form for the joint instantaneous frequency $\omega_\bx(t)$ in terms of rates of change of the ellipse parameters was found by
\citet{lilly11-itsp}, building on work by \citet{lilly10-itsp} for the bivariate case.   In what follows, rates of change of the ellipse angles will be written $\omega_\phi(t)\equiv\phi'(t)$,  $\omega_\theta(t)\equiv\theta'(t)$,  $\omega_\alpha(t)\equiv\alpha'(t)$, and $\omega_\beta(t)\equiv\beta'(t)$ in order to emphasize their interpretation as frequencies.  The instantaneous frequency $\omega_\bx(t)$ then consists of two terms
\begin{equation}
\omega_\bx(t) \label{trivariatefrequency} =\omega_\phi(t)+\sqrt{1-\lambda^2(t)}\,\left[ \omega_\theta(t)+\omega_\alpha(t)\cos\beta(t)\right]
\end{equation}
the first being the rate of change of phase as the particle orbits the ellipse, and the second a {\em generalized precession} which includes the rotation of the ellipse within its plane, as well as azimuthal motion of the plane itself.    If we associate $\omega_\alpha(t)$ with the vertical direction, then $\omega_\alpha(t)\cos\beta(t)$ is the component of the azimuthal motion or ``spinning'' of the plane containing the ellipse that is perpendicular to that plane.  Inserting this into (\ref{frequencyandcirc}) we obtain
\begin{equation}\label{vorticitygeometric}
\frac{1}{2\pi} \,\Gamma_{\!\bx}(t) = \frac{a^2(t)+b^2(t)}{2} \,\omega_\phi(t)  + a(t)b(t)\left[ \omega_\theta(t)+\omega_\alpha(t)\cos\beta(t)\right]
\end{equation}
for the circulation of the canonical ellipse in terms of the rates of change of the ellipse parameters.   It is important to point out that while this has been derived within the context of the canonical ellipse, its form remains valid for any set of ellipse parameters, and thus (\ref{vorticitygeometric}) is the circulation for any flow of the form (\ref{ellipsemodel}).

\subsection{The normal vector and angular momentum}

Next we examine the relationship between the angular momentum vector $\angmom_\bx(t)$ and the normal vector $\bn_\bx(t)$ to the plane instantaneously containing the canonical ellipse.  The latter is given in terms of the ellipse parameters as \citep{lilly11-itsp}
\begin{equation}\label{definenormalvector}
\bn_\bx(t)\equiv\frac{1}{2}\Imag\left\{\bx_+(t) \times{\bx_+^*}(t)\right\}= a(t) b(t)\bJ_3(\alpha(t))\bJ_1(\beta(t))\bk
\end{equation}
where $\bk$ is the vertical unit vector; consequently $\pi\|\bn_\bx(t)\|$ gives the ellipse area.  At each moment, $\angmom_\bx(t)$ is split into a part that is parallel to the normal vector $\bn_\bx(t)$, and a part that is perpendicular to this vector,\begin{equation}
\angmom_\bx(t)=\angmom_\parallel(t)+\angmom_\perp(t)\equiv\left[ \widehat\bn_\bx^T(t)\angmom(t)\right]\widehat\bn_\bx(t) +\left[\angmom(t)-\angmom_\parallel(t)\right]
\end{equation}
where $\widehat \bn_\bx(t)$ is a unit-length version of the normal vector, $\widehat \bn_\bx(t)\equiv  \bn_\bx(t)/\| \bn_\bx(t)\|$.  With straightforward but tedious algebraic manipulations, simple expressions for the angular momentum in terms of the ellipse parameters may be found, as is accomplished in \ref{appendix:angmom}.  The parallel component is given by
\begin{equation}
\label{angmomparallel}
\frac{\angmom_\parallel(t)}{\|\bx_+(t)\|^2}=
\widehat\bn_\bx
\left[\omega_\theta(t)+\omega_\alpha(t) \cos\beta(t)+\sqrt{1-\lambda^2(t)}\,\omega_\phi(t)\right]
\end{equation}
while the perpendicular component is found to be
\begin{multline}\label{angmomperp}
\frac{\angmom_\perp(t)}{\|\bx_+(t)\|^2}
=\frac{1}{2}\,\bJ_3(\alpha(t))\, \bJ_1(\beta(t))\, \bJ_3(\pi/2)
\\ \left\{\bI+\lambda(t)\begin{bmatrix}
\cos\left(2\theta(t)\right)  & \sin\left(2\theta(t)\right) &0 \\ \sin\left(2\theta(t)\right)  & -\cos\left(2\theta(t)\right) &0 \\ 0 & 0 & 0
\end{bmatrix}\right\}
\begin{bmatrix}
\omega_\alpha(t)\sin\beta(t)\\-\omega_\beta(t) \\ 0
\end{bmatrix}
\end{multline}
normalizing both quantities by the magnitude of the analytic vector $\|\bx_+(t)\|^2$ for convenience.  These expressions give the exact instantaneous angular momentum of a ring of particles evolving according to the general kinematic model (\ref{trivariateanalyticfast}).  

We see that the two vectors $\angmom_\bx(t)$ and $\bn_\bx(t)$ specify the plane instantaneously containing the motion in two different ways, the former involving a time derivative of the signal, and the latter the signal's Hilbert transform.  The angular momentum $\angmom_\bx(t)$ and the normal vector $\bn_\bx(t)$ are parallel---and thus agree about the plane containing the ellipse---when the ellipse lies in a fixed plane, but not in general.  An ellipse in a plane that is variable, as specified by a changing normal vector, will generally have a component of the angular momentum vector lying instantaneously within that plane. As seen in (\ref{angmomperp}), the component of the angular momentum lying within the plane containing the ellipse is controlled by the frequencies $\omega_\alpha(t)$ and $\omega_\beta(t) $ that indicate motion of the plane containing the ellipse.

The normal vector is connected in another way to the  angular momentum vector and also to the  tensor-valued first-order moment $\bm{\calJ}_{\!\bx}(t)$.  One may readily show
\begin{align}
\angmom_\bx(t)+\ri\frac{1}{2}\,\bn_\bx'(t)&=\frac{1}{2}\,\bx_+(t) \times{\bx_+'}\!\!^*(t)\\
 \bm{\calJ}_{\!\bx}(t)+\frac{1}{2}\,\cross\,\bn_\bx'(t)&=\frac{1}{2} \Imag \left\{\left[\cross\,\bx_+(t)\right]^H \cross\,\bx_+'(t)\right\}
\end{align}
where in the latter, the right-hand-side is recognized as an expansion of the left-hand-side into a Hermitian and a skew-Hermitian portion. 
Thus the phase-averaged angular momentum $\angmom_\bx(t)$ and the rate of change of normal vector $\bn_\bx'(t)$ are related as real and imaginary parts of the same complex-valued cross-product.  Similarly the matrix $\bm{\calJ}_{\!\bx}(t)$, and the cross-product matrix associated with the rate of change of the normal vector, are related as the Hermitian and skew-Hermitian portions of the same matrix.  In both cases, the term involving $\bn_\bx'(t)$ does not appear in the instantaneous moment $\bS_1(t)$ and thus  does not contribute to the spectral matrix. 

The parallel part of phase-averaged angular momentum vector  $\angmom_\bx(t)$ also appears to be closely related to the instantaneous frequency.    The parallel component of the angular momentum (\ref{angmomparallel}) is very similar to the form the trivariate instantaneous frequency, (\ref{trivariatefrequency}), except that the coefficients of the orbital frequency $\omega_\phi(t)$ and the generalized precession $\omega_\theta(t)+\omega_\alpha(t) \cos\beta(t)$ have been swapped.  This exchange between the phase ``orientation'' specified by $\phi(t)$, and the physical orientation, has been discussed in a very different context.  In examining the general solutions to a freely-evolving two-dimensional shallow water elliptical vortex flow, \citet[][p~397]{holm91-jfm} notes that reversing the roles of the orientation angle and phase angle also exchanges the vortex circulation and its momentum, an effect he refers to as the ``Dedekind duality".   This comment in fact proved to be an essential clue pointing towards the unity of instantaneous frequency and circulation as expressed by (\ref{frequencyandcirc}).

\subsection{Bandwidth, angular momentum, and kinetic energy}

Finally the relationship between bandwidth, angular momentum, and kinetic energy is examined.    We will shortly show that these three quantities are related by
\begin{multline}\label{wewillshow}
 \frac{\frac{1}{2} \overline{\,\left\|\bx'(t)\right\|^2}^{\,\varphi}}{\overline{\,\|\bx(t)\|^2}^{\,\varphi}} =\frac{1}{2}\,\left|\frac{\Gamma_\bx(t)}{2\pi \overline{\,\left\|\bx(t)\right\|^2}^{\,\varphi}}\right|^2+\frac{1}{2}\upsilon_\bx^2(t) \\= 
  \overset{(\ri)}{\overbrace{\frac{1}{2}\,\left|\frac{\Gamma_\bx(t)}{2\pi \overline{\,\left\|\bx(t)\right\|^2}^{\,\varphi}}\right|^2 }}+\overset{(\ri\ri)}{\overbrace{\frac{1}{2}\,\left|\frac{\kappa_\bx'(t)}{\kappa_\bx(t)}\right|^2}} + \overset{(\ri\ri\ri)}{\overbrace{\frac{\overline{\,\left\|\bx(t)\times\bx'(t)-\overline{\bx(t)\times\bx'(t)}^{\,\varphi} \right\|^2}^{\,\varphi} }{\left|\overline{\,\left\|\bx(t)\right\|^2}^{\,\varphi}\right|^2}}}
\end{multline}
and thus the phase-averaged kinetic energy of the canonical ellipse can be partitioned into three distinct portions: (i) the kinetic energy associated with the circulation; (ii) kinetic energy due to expansion and contraction of the ellipse; and (iii) kinetic energy associated with deviation of the  angular momentum vector from its mean value around the ellipse periphery---the {\em angular momentum variance}.  This result remains valid for any flow of the general kinematic form (\ref{ellipsemodel}).  For the canonical ellipse, it is also seen that one-half the squared instantaneous bandwidth can be interpreted as an {\em excess kinetic energy} above that associated with the circulation. 

In \citet{lilly11-itsp}, the squared instantaneous bandwidth is found to contain four non-negative terms involving various rates of change of the ellipse geometry,
\begin{multline}
\upsilon_\bx^2(t)=\label{alternatebandwidth}
\left|\frac{\kappa_\bx'(t)}{\kappa_\bx(t)}\right|^ 2\\+
\frac{1}{4}\frac{\left|\lambda'(t)\right|^2 }{1- \lambda^2(t)}+ \lambda^2(t)\left[ \omega_\theta(t)+\omega_\alpha(t)\cos\beta(t)\right]^2
+\frac{\left|\bx_+^H(t)\widehat\bn_\bx'(t)\right|^2}{\left\| \bx_+(t)\right\|^2}.
\end{multline}
From left to right, these represent the effects of the rate of amplitude modulation $\kappa_\bx'(t)$, the deformation rate $\lambda'(t)$, and the generalized precession, together with a fourth term that is associated entirely with three-dimensional effects due to 
the time variation of the orientation of the plane containing the ellipse.   Comparison of (\ref{alternatebandwidth}) with (\ref{wewillshow}) shows that all terms except the amplitude modulation term are associated with the angular momentum variance.

To prove (\ref{wewillshow}), we first note that the variance of a cross product along the canonical ellipse is given by, as
shown in \ref{appendix:phaseaveraging},
\begin{equation}\label{crossproducttheorem}
\overline{\,\left\|\Real\mathbf{f}(t)\times\Real\bg(t)-\overline{\Real\mathbf{f}(t)\times\Real\bg(t)}^{\,\varphi}\right\|^2}^{\,\varphi}
=\frac{1}{8}\left\|\mathbf{f}(t)\times \bg(t)\right\|^2
\end{equation}
where $\mathbf{f}(t)$ and $\bg(t)$ are analytic vectors equal to some constant times $\bx_+(t)$ or one of its time derivatives; note carefully that $\bg(t)$ on the RHS is not conjugated.   The phase variance of the angular momentum can therefore be expressed as
\begin{equation}\label{crossproduct}
\overline{\,\left\|\bx(t)\times\bx'(t)-\overline{\bx(t)\times\bx'(t)}^{\,\varphi}\right\|^2}^{\,\varphi}
=\frac{1}{8}\left\|\bx_+(t)\times \bx_+'(t)\right\|^2
\end{equation}
in terms of the analytic signal $\bx_+(t)$ and its first derivative. Lagrange's identity for complex-valued vectors is, as discussed in \ref{appendix:phaseaveraging},
\begin{equation}
\|\mathbf{f}\times\bg\|^2= \|\mathbf{f}\|^2\|\bg\|^2- |\bg^H\mathbf{f}|^2\label{lagrange}
\end{equation}
and consequently the cross-product term in (\ref{crossproduct}) becomes
\begin{equation}\label{lagrangemomentum}
\frac{\|\bx_+(t)\times\bx_+'(t)\|^2}{\left\|\bx_+(t)\right\|^4} =
 \frac{\|\bx_+'(t)\|^2}{\left\|\bx_+(t)\right\|^2} -
 \left|\frac{\kappa_\bx'(t)}{\kappa_\bx(t)}\right|^2
-\omega_\bx^2(t)
=  \upsilon_\bx^2(t) - \left|\frac{\kappa_\bx'(t)}{\kappa_\bx(t)}\right|^2
\end{equation}
after making use of the definition of the joint instantaneous bandwidth (\ref{bandwidth}).   Combining (\ref{lagrangemomentum}) with (\ref{crossproduct}) then gives (\ref{wewillshow}).

\section{Conclusions}\label{section:conclusions}

The purpose of this paper has been to introduce and interpret matrix-valued generalizations, appropriate to the analysis of modulated oscillations in two or three dimensions, of the signal processing concepts of instantaneous amplitude, frequency, and bandwidth.  These matrix-valued instantaneous moments are found to be identical to the physical moments of the canonical ellipse that is assigned to the signal by the action of taking its analytic part.   This gives an illuminating new interpretation of the spectral matrix as being composed of time-varying contributions from the physical moments of the canonical ellipse.   Since the spectral matrix is fundamental to data analysis in many fields, these ideas to informatively decompose its variability across time could potentially find widespread application.   A new result partitioning the kinetic energy of an elliptical vortex into three portions---associated with the circulation, isotropic expansion and contraction, and the {\em variance} of the angular momentum around the ellipse periphery---was also presented.

A direct connection to a recently developed method for treating modulated oscillations in real-world data should be mentioned.    In practice, oscillatory signals exist not in isolation but juxtaposed with noise or other signal variability.  One is therefore faced with the problem of extracting the presumed oscillatory signals from a noisy background.  A powerful solution to this problem is {\em wavelet ridge analysis} \citep{delprat92-itit,mallat,lilly10-itit}, which is based upon the use of an analytic wavelet transform to isolate the oscillatory signal from surrounding variability, leading to estimates of the analytic signal and associated instantaneous moments.    Recently this method was extended to the multivariate case by \citet{lilly12-itsp}, who also solve for the deterministic error terms arising from non-negligible modulation.  The results of this multivariate wavelet ridge analysis can be used to give {\em estimates} of the instantaneous moment matrices corresponding to an oscillatory signal isolated from the surrounding variability, to which the ideas in the present paper could then be  immediately applied.

\begin{acknowledgements}
This work was support by grant \#1031002 from the Physical Oceanography program of the United States National Science Foundation.
\end{acknowledgements}

\appendix{A quartic phase average}\label{appendix:phaseaveraging}

In this appendix we prove a phase averaging result for the quartic ($M=4$) case, similar to the quadratic results (\ref{realimagtheorem}) and (\ref{realimagtheorem2}). Observe from (\ref{abouttoaverage}) that the phase average of a product of four real parts, with $\xi_n(t)\equiv e^{\ri \phi(t)}[c_n(t)+\ri d_n(t)]$, is
\begin{multline}\label{almostproved}
\overline{\Real\xi_1(t)\Real\xi_2(t) \Real\xi_3(t)\Real\xi_4(t)}^{\,\varphi} \\=
\frac{1}{4}\left[c_1(t)c_2(t)+d_1(t)d_2(t)\right] \left[c_3(t)c_4(t)+d_3(t)d_4(t)\right] \\
+\frac{1}{8}\left[c_1(t)c_2(t)-d_1(t)d_2(t)\right] \left[c_3(t)c_4(t)-d_3(t)d_4(t)\right]\\+
\frac{1}{8}\left[c_1(t)d_2(t)+c_2(t)d_1(t)\right] \left[c_3(t)d_4(t)+c_4(t)d_3(t)\right]
\end{multline}
since all other terms involve sinusoidal functions of $\phi(t)$ that vanish upon the phase integration.  Comparing the last two terms on the RHS with
\begin{multline}
\Real\left\{\xi_1(t)\xi_2(t)\xi_3^*(t)\xi_4^*(t)\right\}\\=
\left[c_1(t)c_2(t)-d_1(t)d_2(t)\right] \left[c_3(t)c_4(t)-d_3(t)d_4(t)\right]+\\
\left[c_1(t)d_2(t)+c_2(t)d_1(t)\right] \left[c_3(t)d_4(t)+c_4(t)d_3(t)\right]
\end{multline}
and comparing the first term on the RHS of (\ref{almostproved}) with (\ref{realimagtheorem}),
we obtain
\begin{multline}\label{phasevarianceresult}
\overline{\,\Real\xi_1(t)\Real\xi_2(t)\Real\xi_3(t)\Real\xi_4(t)}^{\,\varphi} =\\
\overline{\Real\xi_1(t)\Real\xi_2(t)}^{\,\varphi}\,\overline{\Real\xi_3(t)\Real\xi_4(t)}^{\,\varphi}+
\frac{1}{8}\Real\left\{\xi_1(t)\xi_2(t)\xi_3^*(t)\xi_4^*(t)\right\}.
\end{multline}
A special case occurs for $\xi_3(t)=\xi_1(t)$ and $\xi_4(t)=\xi_2(t)$, in which case  (\ref{phasevarianceresult}) becomes, after some rearranging,
\begin{equation}\label{phasevariance}
\overline{\,\left|\Real\xi_1(t)\Real\xi_2(t)-\overline{\Real\xi_1(t)\Real\xi_2(t)}^{\,\varphi}\right|^2}^{\,\varphi} = \frac{1}{8}\left|\xi_1(t)\right|^2\left|\xi_2(t)\right|^2 
\end{equation}
which gives a simple expressions for the phase variance of $\Real\xi_1(t)\Real\xi_2(t)$ in terms of $\xi_1(t)$ and $\xi_2(t)$. Comparison with  (\ref{realimagtheorem}) shows this can then be related to the product of the phase averages of the squares of $\Real\xi_1(t)$ and $\Real\xi_2(t)$.

Next we prove Lagrange's identity, presented later in (\ref{lagrange}), for two complex-valued vectors $\mathbf{f}$ and~$\bg$.  This can be proven using standard vector identities \citep[see e.g.][p. 4]{lindell}, but here we will write out components for future reference.  The squared norm of the cross-product of $\mathbf{f}$ and~$\bg$ is given by
\begin{multline}
\|\mathbf{f}\times\bg\|^2=|f_yg_z-f_zg_y|^2+|f_xg_z-f_zg_x|^2+|f_xg_y-f_yg_x|^2\\=
|f_y|^2|g_z|^2+|f_z|^2|g_y|^2+
|f_x|^2|g_z|^2+|f_z|^2|g_x|^2+
|f_x|^2|g_y|^2+|f_y|^2|g_x|^2\\
-2\Real\{f_yf_z^*g_y^*g_z\}
-2\Real\{f_xf_z^*g_x^*g_z\}
-2\Real\{f_xf_y^*g_x^*g_y\}\label{almostlagrange}
\end{multline}
while the squared magnitude of the Hermitian inner product $\bg^H\mathbf{f}$ is
\begin{multline}
 |\bg^H\mathbf{f}|^2=|f_xg_x^*+f_yg_y^*+f_zg_z^*|^2
 =|f_x|^2|g_x|^2+|f_y|^2|g_y|^2+|f_z|^2|g_z|^2\\
 +2\Real\left\{f_yf_z^*g_y^*g_z\right\}
 +2\Real\left\{f_xf_z^*g_x^*g_z\right\}
 +2\Real\left\{f_xf_y^*g_x^*g_y\right\}.
\end{multline}
Comparison of these two equations proves Lagrange's identity (\ref{lagrange}).

Finally we prove (\ref{crossproducttheorem}) for the phase variance of a cross product.   The real-valued cross product $\Real\mathbf{f}(t)\times\Real\bg(t)$ has a $z$ component equal to
\begin{multline}
\left[\Real\mathbf{f}(t)\times\Real\bg(t)\right]_z =
\left[\Real f_x(t) \Real g_y(t)-\overline{\Real f_x(t) \Real g_y(t)}^{\,\varphi}\right] \\ -\left[\Real f_y(t) \Real g_x(t)-\overline{\Real f_y(t) \Real g_x(t)}^{\,\varphi}\right]
\end{multline}
the square of which will contain three terms, two squares and a cross term.  The phase averages of the first two terms are
\begin{align}
\overline{\,\left|\Real f_x(t) \Real g_y(t)-\overline{\Real f_x(t) \Real g_y(t)}^{\,\varphi}\right|^2}^{\,\varphi}& = \frac{1}{8} |f_x(t)|^2|g_y(t)|^2\\
\overline{\,\left|\Real f_y(t) \Real g_x(t)-\overline{\Real f_y(t) \Real g_x(t)}^{\,\varphi}\right|^2}^{\,\varphi}&= \frac{1}{8} |f_y(t)|^2|g_x(t)|^2
 \end{align}
which have simplified on account of the phase variance relation (\ref{phasevariance}).  The phase average of the cross term is $(-2)$ times
\begin{multline}
\overline{\left[\Real f_x(t) \Real g_y(t)
-\overline{\Real f_x(t) \Real g_y(t)}^{\,\varphi}\right]\left[\Real f_y(t) \Real g_x(t)-\overline{\Real f_y(t) \Real g_x(t)}^{\,\varphi}\right]}^{\,\varphi}\\
=\overline{\,\Real f_x(t) \Real g_y(t)\Real f_y(t) \Real g_x(t)}^{\,\varphi}
- \overline{\Real f_x(t) \Real g_y(t)}^{\,\varphi}\,\overline{\Real f_y(t) \Real g_x(t)}^{\,\varphi}\\
=\frac{1}{8}\Real\left\{ f_x(t) g_y(t)  f_y^*(t)  g_x^*(t)\right\}
\end{multline}
using the quartic phase averaging result (\ref{phasevarianceresult}) to obtain the final line.  Combining the previous three results we find
\begin{multline}
\overline{\left|\left[\Real\mathbf{f}(t)\times\Real\bg(t)\right]_z-\overline{\left[\Real\mathbf{f}(t)\times\Real\bg(t)\right]_z}^{\,\varphi} \right|}^{\,\varphi}
\\= \frac{1}{8}\left[ |f_x(t)|^2|g_y(t)|^2 +|f_y(t)|^2|g_x(t)|^2 - 2 \Real\left\{ f_x(t) g_y(t)  f_y^*(t)  g_x^*(t)\right\}  \right]
\end{multline}
but observe that this expression occurs as one of three contributions to $\|\mathbf{f}\times\bg\|^2$ in (\ref{almostlagrange}).     Considering the similar contributions from the $x$ and $y$ components of  $\Real\mathbf{f}(t)\times\Real\bg(t)$ leads to (\ref{crossproducttheorem}) for the phase variance of a cross product.

\appendix{Angular momentum derivation}\label{appendix:angmom}

In this appendix an expression is derived for the phase-averaged angular momentum in terms of the ellipse parameters.   Here we introduce the 2-vector
\begin{equation}\widetilde\bx_+(t)
=\re^{\ri\phi(t)}
\bJ(\theta(t))\,
   \br(t)
   \label{twodellipse} =   \re^{\ri\phi(t)}  \begin{bmatrix}
   \cos \theta(t) & -\sin\theta(t)\\
   \sin \theta(t)  & \cos\theta(t)
   \end{bmatrix} \begin{bmatrix}
   a(t) \\-\ri b(t)
   \end{bmatrix}
\end{equation}
where $\bJ(\theta)$ is the $2\times2$ counterclockwise rotation matrix and $\br(t)\equiv\begin{bmatrix}a(t)\! \!&\!-\ri b(t)
   \end{bmatrix}^T$.  The rate of change of this vector is, letting $\bJ$ with no argument indicate the ninety-degree rotation matrix $\bJ\equiv\bJ(\pi/2)$, and with $\lambda(t) \equiv  \frac{a^2(t)-b^2(t)}{a^2(t)+b^2(t)}$, 
\begin{multline}\label{xtildederivdef2}
\widetilde\bx_+'(t)=e^{\ri\phi(t)}\bJ(\theta(t)) \left\{\left[\frac{\kappa_\bx'(t)}{\kappa_\bx(t)} +\ri\omega_\phi(t)
+\ri\sqrt{1-\lambda^2(t)}\,\omega_\theta(t)
\right]\br(t) \right.\\\left.
+ \left[\lambda(t) \omega_\theta(t)+ \ri\frac{1}{2}\frac{\lambda'(t)}{\sqrt{1-\lambda^2(t)}}\right]\bJ\br^*(t)
\right\}
\end{multline}
as may be derived in a few lines of algebra. Alternatively (\ref{xtildederivdef2}) follows from (69) of Appendix~B of \citet{lilly11-itsp} if one notes $\bJ\br(t) = \ri \sqrt{1-\lambda^2(t)}\,\br(t) +\lambda(t) \bJ \br(t)$.  Recall that in  (\ref{xtildederivdef2}),  $\omega_\phi(t)\equiv\phi'(t)$ and $\omega_\theta(t)\equiv\theta'(t)$ as defined in the text.

The complex-valued cross product $\bx_+(t)\times{\bx_+^*}\!'(t)$ can be now be conveniently evaluated.  The three-vector $\bx_+(t)$ defined in (\ref{trivariateanalytic})  becomes
\begin{equation}
\bx_+(t)=\bJ_3(\alpha(t))\, \bJ_1(\beta(t))\bH\,\widetilde\bx_+(t)
\end{equation}
in terms of $\widetilde\bx_+(t)$,  where we have introduced
\begin{equation}
\bH=\begin{bmatrix}1&0\\0&1\\0 &0\end{bmatrix}
\end{equation}
as the $3\times 2$ matrix which maps a 2-vector onto the horizontal plane in three dimensions.   The rate of change of $\bx_+(t)$ is
\begin{equation}
\bx_+'(t)=\label{derivpart1}
\bJ_3(\alpha(t))\, \bJ_1(\beta(t))\bH\,\widetilde\bx_+'(t)+\left[\bJ_3(\alpha(t))\, \bJ_1(\beta(t))\right]'\bH\,\widetilde\bx_+(t)
\end{equation}
where the latter is found to take the form
\begin{multline}\label{derivpart2}
\left[\bJ_3(\alpha(t))\, \bJ_1(\beta(t))\right]'\bH\,\widetilde\bx_+(t)=\\ \left[\omega_\alpha(t)\cos\beta(t)\right]\bJ_3(\alpha(t))\bJ_1(\beta(t))\bH
\bJ\widetilde\bx_+(t)
-\left\{\bx_+^T(t)\widehat\bn_\bx'(t)\right\}\widehat\bn_\bx(t).
\end{multline}
Observe that the real and imaginary parts of the term involving $\widetilde\bx_+(t)$ lie within the plane of the ellipse, whereas the final term is collinear with the normal vector.

With $\bR$ a real orthogonal matrix such that $\bR^T=\bR^{-1}$, and having unit determinant so that $\bR$ is a proper rotation matrix, the cross product transforms as $\left(\bR \mathbf{f}\right)\times \left(\bR\bg\right)=\bR\left( \mathbf{f}\times \bg\right)$ for two complex-valued vectors $ \mathbf{f}$ and  $\bg$.  This may be proven for real-valued vectors by standard vector identities, with the complex-valued case following by writing out the real and imaginary parts to obtain a set of four real-valued cross products.   Also note that for two 2-vectors $\mathbf{a}$ and $\mathbf{b}$, we have
\begin{equation}
\left(\bH\ba\right)\times\left(\bH\bb^*\right) =\left[\bb^H\bJ\ba\right]\bk
\end{equation}
as may be readily verified.  We then find the following expressions
\begin{align}
\bH\br(t)\times\bH\br^*(t) & = \left[\br^H(t)\bJ\br(t)\right]\bk 
= \ri \sqrt{1-\lambda^2(t)}\,\|\bx_+(t)\|^{2}\,\bk\\
\bH\br(t)\times\bH\bJ\br(t) & = \left[\br^T(t)\br(t)\right]\bk = \lambda(t) \|\bx_+(t)\|^{2}\,\bk
\end{align}
for the cross-product between $\bH\br(t)$ and its own conjugate or itself rotated by ninety degrees.   Combining these with (\ref{derivpart1}) and (\ref{derivpart2}), we find
\begin{multline}
\frac{\left[\bx_+(t) \times {\bx_+'}\!\!^*(t)\right]_\parallel}{\|\bx_+(t)\|^2}  =
\left\{\sqrt{1-\lambda^2}\left[\omega_\phi(t)-\ri\frac{\kappa_\bx'(t)}{\kappa_\bx(t)}\right] \right.\\\left.
+\ri\frac{1}{2}\frac{\lambda(t)\lambda'(t)}{\sqrt{1-\lambda^2(t)}}+\omega_\theta(t)+\omega_\alpha(t)\cos\beta(t) \right\}\widehat\bn_\bx(t)
\end{multline}
for the parallel part of the complex-valued cross-product $\bx_+(t) \times {\bx_+'}\!\!^*(t)$, i.e. that part parallel to the normal vector $\bn_\bx(t)$.

To find the perpendicular part of the cross-product, we will make use of the fact that $\bH\mathbf{a}\times
\bk = -\bH\bJ\mathbf{a}$ for a complex two-vector $\ba$.  Then
\begin{equation}\label{parallelcross}
\frac{\left[\bx_+(t) \times {\bx_+'}\!\!^*(t)\right]_\perp}{\|\bx_+(t)\|^2}  = \left[ \frac{\bx_+^H(t)\widehat\bn_\bx'(t)}{\|\bx_+(t)\|}\right]\bJ_3(\alpha(t))\bJ_1(\beta(t)) \bH\frac{\bJ\widetilde\bx_+(t)}{\|\bx_+(t)\|}
\end{equation}
gives the parallel part of the cross product.    Observe that this has a magnitude
\begin{equation}
\left\|\frac{\left[\bx_+(t) \times {\bx_+'}\!\!^*(t)\right]_\perp}{\|\bx_+(t)\|^2} \right\|^2 =\frac{ \left| \bx_+^H(t)\widehat\bn_\bx'(t)\right|^2}{\|\bx_+(t)\|^2}
\end{equation}
and a complex direction which lies in the plane of the ellipse, rotated ninety degrees from the complex direction of the analytic signal.   We may rearrange (\ref{parallelcross}) to find
\begin{equation}
\frac{\left[\bx_+(t) \times {\bx_+'}\!\!^*(t)\right]_\perp}{\|\bx_+(t)\|^2}=
\bJ_3(\alpha(t))\bJ_1(\beta(t)) \bH\bJ\,
\frac{\widetilde\bx_+(t)\widetilde\bx_+^H(t)}{\|\bx_+(t)\|^2}\begin{bmatrix}
\omega_\alpha(t)\sin\beta(t)\\-\omega_\beta(t)\label{rearranged}
\end{bmatrix}
\end{equation}
using the definition (\ref{twodellipse}) of $\widetilde\bx_+(t)$ together with 
\begin{equation}\label{nprime}
\widehat\bn_\bx'(t)=\bJ_3(\alpha(t))\bJ_1(\beta(t))\begin{bmatrix}
\omega_\alpha(t)\sin\beta(t) &
-\omega_\beta(t)&0
\end{bmatrix}^T
\end{equation}
for the rate of change of the unit normal vector.  The $2\times 2$ matrix occurring in (\ref{rearranged}) has a real part
\begin{multline}\label{almostdoneparallel}
\frac{\Real\left\{\widetilde\bx_+(t)\widetilde\bx_+^H(t)\right\}}{\|\bx_+(t)\|^2}=
\frac{1}{a^2(t)+b^2(t)}\,\bJ(\theta(t))
\begin{bmatrix}
 a^2(t) & 0 \\0 & b^2(t)
\end{bmatrix}\bJ^T(\theta(t)) \\
= \frac{1}{2}\bI_2+\frac{1}{2}\lambda(t)\begin{bmatrix}
\cos\left(2\theta(t)\right)  & \sin\left(2\theta(t)\right) \\ \sin\left(2\theta(t)\right)  & -\cos\left(2\theta(t)\right)
\end{bmatrix}
\end{multline}
where $\bI_2$ is the $2\times 2$ identity matrix.   We then obtain (\ref{angmomperp}) for the perpendicular part of the angular momentum vector from (\ref{rearranged}) by noting $\bH\bJ=\bJ_3(\pi/2)\bH$.

\label{lastpage}

\end{document}

%% file: texheader.tex
\newcommand{\tr}{\ensuremath{\mathrm{tr}}}

\newcommand{\bzero}{\ensuremath{\mathbf{0}}}

\newcommand{\ba}{\ensuremath{\mathbf{a}}}
\newcommand{\bb}{\ensuremath{\mathbf{b}}}

\newcommand{\bg}{\ensuremath{\mathbf{g}}}

\newcommand{\bj}{\ensuremath{\mathbf{j}}}
\newcommand{\bk}{\ensuremath{\mathbf{k}}}

\newcommand{\bn}{\ensuremath{\mathbf{n}}}

\newcommand{\br}{\ensuremath{\mathbf{r}}}

\newcommand{\bu}{\ensuremath{\mathbf{u}}}

\newcommand{\bx}{\ensuremath{\mathbf{x}}}

\newcommand{\bz}{\ensuremath{\mathbf{z}}}

\newcommand{\bF}{\ensuremath{\mathbf{F}}}

\newcommand{\bH}{\ensuremath{\mathbf{H}}}
\newcommand{\bI}{\ensuremath{\mathbf{I}}}
\newcommand{\bJ}{\ensuremath{\mathbf{J}}}

\newcommand{\bQ}{\ensuremath{\mathbf{Q}}}
\newcommand{\bR}{\ensuremath{\mathbf{R}}}
\newcommand{\bS}{\ensuremath{\mathbf{S}}}

\newcommand{\bX}{\ensuremath{\mathbf{X}}}

\newcommand{\calA}{\ensuremath{\mathcal{A}}}

\newcommand{\calE}{\ensuremath{\mathcal{E}}}

\newcommand{\calH}{\ensuremath{\mathcal{H}}}
\newcommand{\calI}{\ensuremath{\mathcal{I}}}
\newcommand{\calJ}{\ensuremath{\mathcal{J}}}

